\documentclass[journal]{IEEEtran}

\usepackage{amsmath,amssymb,amsfonts}
\usepackage{booktabs}
\usepackage{cite}
\usepackage{graphicx}
\usepackage[caption=false,font=footnotesize]{subfig}
\usepackage{xcolor}

\begin{document}
	
	\title{DeepRT-Cal: Learning to Calibrate Ray-Traced Twin Channel Statistics with Low-Fidelity Environmental Geometry}
	
	\author{Shaoyi~Liu,~\IEEEmembership{Student Member,~IEEE},
		Jianhua~Zhang,~\IEEEmembership{Fellow,~IEEE},
		Li~Yu,~\IEEEmembership{Member,~IEEE},
		Yichen~Cai,~\IEEEmembership{Student Member,~IEEE},
		Tao~Wu,~\IEEEmembership{Student Member,~IEEE},
		Yuxiang~Zhang,~\IEEEmembership{Member,~IEEE},
		Jikun~Du,~\IEEEmembership{Student Member,~IEEE},
		and~Guangyi~Liu,~\IEEEmembership{Senior Member,~IEEE},

        \thanks{ This work was supported in part by the Guangdong Major Project of Basic and Applied Basic Research (No. 2023B0303000001); in part by the National Natural Science Foundation of China under Grant 62401084 and Grant 62525101; and the Beijing University of Posts and Telecommunications-China Mobile Communications Group Co., Ltd. Joint Institute.}
		\thanks{Corresponding author (email: jhzhang@bupt.edu.cn; li.yu@bupt.edu.cn)} 
		\thanks{S. Liu, J. Zhang, L. Yu, Y. Cai, T. Wu, Y. Zhang, and J. Du are with the State Key Laboratory
			of Networking and Switching Technology, Beijing University of Posts and
			Telecommunications, Beijing 100876, China. G. Liu is with the China Mobile Research Institute, Beijing 100053, China}

		}
	\markboth{IEEE Transactions on Antennas and Propagation,~Vol.~XX, No.~XX, 2026}%
	{Author \MakeLowercase{\textit{et al.}}: Geometry-Aware Calibration of Ray-Traced Channel Statistics}
	
	\maketitle
	
	\begin{abstract}
		Sixth-generation (6G) wireless systems increasingly rely on digital twin channels (DTCs) to represent and interact with physical radio environments. Ray tracing (RT) is naturally suited to DTC construction because it provides deterministic and physically interpretable channel predictions. However, high-fidelity three-dimensional environment models are expensive to obtain, whereas geometric inaccuracies in low-fidelity models can alter the detected multipath components (MPCs) and consequently distort received-power and delay-domain twin channel statistics. 
        Building on the DeepRT ray-tracing framework, this paper proposes DeepRT-Cal, a geometry-aware cross-fidelity calibration framework that predicts high-fidelity twin channel statistics from low-fidelity geometry and RT features.
		The network encodes global link descriptors, masked low-fidelity MPC tokens, propagation-corridor geometry, and link-centered environmental objects. During training, detailed geometry is introduced as privileged information with structured masking, while deployment requires only low-fidelity geometry and low-fidelity RT information. A link-conditioned cross-attention module retrieves the environmental context relevant to each Tx--Rx link. The prediction heads estimate high-fidelity twin channel statistics, including total received power, RMS delay spread, mean excess delay, and a normalized power-weighted power-delay profile (PDP). Outdoor evaluation shows that, in an non-line of sight (NLoS)  region, the method reduces power MAE from 16.80 to 3.88 dB and RMS delay-spread MAE from 304.89 to 98.19 ns. In a LoS-dominated region, power prediction improves, while RMS delay-spread accuracy remains comparable to direct RT.
	\end{abstract}
	
	\begin{IEEEkeywords}
		Channel modeling, ray tracing, digital twin channel, cross-attention, wireless propagation.
	\end{IEEEkeywords}
	
	\section{Introduction}

	With the rapid evolution of the sixth-generation (6G) mobile communication networks, the digital twin channel (DTC) has emerged as a crucial enabling technology to bridge the physical and digital electromagnetic worlds \cite{3_wanghengDTC, EIC, 6_zhang2023channel, 5_NVIDADTN, REKP}. To realize a high-fidelity DTC, it is imperative to precisely model the complex interactions between propagating waves and the physical environment. Recent efforts toward Third Generation Partnership Project (3GPP) 6G standardization have also addressed integrated sensing and communication (ISAC) channel modeling~\cite{11359163} and radar cross section (RCS) modeling of typical targets~\cite{11159279}, further motivating accurate modeling of radio interactions with the physical environment. In this context, ray tracing (RT) naturally aligns with the stringent requirements of DTCs. Relying on high-frequency asymptotic techniques, RT provides a deterministic computational framework that accurately replicates underlying electromagnetic propagation mechanisms, offering physically interpretable predictions and deterministic insights into wireless channels \cite{5_NVIDADTN, 2_DTN2024WirelessCommunications, Lu2023}.

	The deployment of RT in 6G DTCs introduces two primary demands: computational speed and predictive precision. Aided by the exponential growth in modern computing power (e.g., graphics processing unit (GPU) acceleration) and the continuous proposal of novel RT acceleration algorithms, the historical bottleneck of computational speed has been progressively alleviated \cite{4_beijiaoRT,  EA_SBR, miao2024enhanced, DRT1, DRT2, DRT3}. However, the precision issue of RT remains largely unresolved. The accuracy of deterministic RT is heavily bounded by the fidelity of its input environment. One major source of error originates from the inaccuracy of electromagnetic material properties. Slight deviations in dielectric constant or conductivity can lead to significant distortions in channel feature calculations (e.g., reflection loss), and acquiring material parameters that perfectly match the real world remains highly challenging \cite{OpenGERT, Rizk2000, 10_NYURay}. Another, and perhaps more severe, source of error is the geometric discrepancy in three-dimensional (3D) environment modeling. Currently, obtaining high-precision digital twin models (such as millimeter-level light detection and ranging (LiDAR) point clouds) incurs prohibitive costs and is impractical for large-scale ubiquitous deployment \cite{liuISAC, Liu2023}. Conversely, when utilizing cost-effective, low-precision 3D maps (e.g., open-source OpenStreetMap (OSM) data), inherent topological errors---such as over-smoothed surfaces, displaced building boundaries, or missing fine structures---severely compromise the RT pathfinding process, which is strongly constrained by the geometric propagation mechanisms implemented by the solver \cite{15_yuanRTMIMO}. These errors are not limited to small perturbations in path loss. They can change whether a path is visible, move the reflection or diffraction point, alter the propagation distance and interaction angle, or change the blockage relation between a transmitter (Tx) and a receiver (Rx). As a result, the low-fidelity solver may omit a valid path, introduce a path that is absent from the detailed scene, or produce a multipath component (MPC) with substantially different delay and power. The resulting low- and high-fidelity MPC sets can therefore differ in cardinality, topology, and delay-power distribution, even for the same Tx--Rx locations and radio settings. These differences propagate to several channel statistics: changes in reflection and blockage affect total received power, while changes in delayed energy affect root-mean-square (RMS) delay spread, mean excess delay, and the complete power-delay profile (PDP). This geometric degradation inevitably leads to the misalignment of multipath trajectories or even the complete loss of valid physical paths, motivating a channel-level cross-fidelity calibration method rather than a path-by-path correction.
	
	In recent years, artificial intelligence (AI) has developed rapidly, demonstrating formidable capabilities in solving complex, high-dimensional non-linear mapping problems. Naturally, researchers have begun exploring the integration of AI with RT to address its inherent limitations. Current literature on AI-assisted RT can be broadly categorized into two paradigms. The first category focuses on constructing ``Surrogate Models'' \cite{DeepRay, EM_DeepRay, Winert, RayProNet}. These studies utilize deep learning architectures, such as convolutional neural networks (CNNs) or generative models, to learn the mapping from environment maps to channel characteristics (e.g., RadioUNet), thereby significantly accelerating the RT computational process \cite{radioUnet ,yu2024channelgpt, zhang2025four, shi2025can}. While these approaches effectively reduce computational overhead, the fundamental nature of surrogate models dictates a critical limitation: the ``student'' (the AI model) inherits the limitations of the ``teacher'' (the RT supervision target). Because these models are supervised entirely by the deterministic outputs of RT, they inherit all the inherent physical and geometric errors of the baseline simulator, offering speed at the expense of true physical fidelity.
	
	The second category leverages the backpropagation characteristics of modern AI frameworks to perform end-to-end calibration of the electromagnetic parameters involved in RT calculations. A prominent representative of this approach is Nvidia's Sionna RT, a differentiable RT platform \cite{sionna_1}. Differentiable RT allows gradients to flow through geometric propagation paths, enabling the reverse optimization of material properties (such as relative permittivity and conductivity) using gradient descent by comparing the simulated channel impulse response (CIR) with real-world measurements \cite{sionna_2}. However, material-parameter calibration can modify the response of paths already generated by the solver, but it cannot by itself restore a propagation path that is absent from the topology induced by the input geometry. If an MPC is entirely lost during the ray-finding phase due to missing topological details (e.g., an over-smoothed edge in a low-precision 3D model), fine-tuning material parameters cannot recover this missing physical path. Existing AI-assisted studies have considered RT acceleration, radio-map estimation, and channel super-resolution~\cite{Zhang2023}, but the compensation of geometry-induced changes in the MPC topology and delay-power statistics remains comparatively underexplored.
	
	To bridge the aforementioned critical gap, this paper proposes the DeepRT-Cal cross-fidelity calibration framework. Unlike existing calibration methods that solely adjust material parameters, DeepRT-Cal takes the low-fidelity 3D environmental topology and its corresponding coarse RT multipath results as joint inputs, employing a cross-attention mechanism to compensate for geometric errors. The ultimate goal is to synthesize high-fidelity channel statistical characteristics that closely approximate those derived from detailed 3D models. The main contributions of this paper are summarized as follows:

	\begin{itemize}
		
		\item Cross-Fidelity RT Formulation and Link-Centered Object Tokenization:
		We formulate low-to-high-fidelity RT enhancement as a channel-level calibration problem. The supervised scalar targets are total received power, RMS delay spread, and mean excess delay, while the power-weighted excess-delay PDP is represented by a normalized 200-bin distribution. Furthermore, we develop a link-centric object tokenization strategy to preserve the environmental structures most relevant to channel formation, while avoiding the computational burden and correspondence ambiguity associated with directly processing complete 3D meshes.
		
		\item Design of DeepRT-Cal with Link-Conditioned Cross-Attention:
		We design DeepRT-Cal, a heterogeneous feature fusion network that separately encodes global link attributes, low-fidelity multipath features, and environmental object representations. By generating a dynamic query from the link-level and multipath features to perform multi-head cross-attention over object tokens, DeepRT-Cal learns the implicit nonlinear relationship between propagation-related evidence, environmental geometry, and high-fidelity channel responses.
		
		\item Performance Evaluation and Ablation Studies in Complex Outdoor Scenarios:
		We conduct extensive numerical experiments in complex outdoor RT scenarios to evaluate the proposed framework under low-fidelity environmental and RT inputs. The results demonstrate that DeepRT-Cal substantially improves the prediction accuracy of key channel metrics and delay-domain power distributions over direct low-fidelity RT outputs, while ablation studies further verify the effectiveness and contribution of the proposed cross-attention module.

	\end{itemize}
	
	\begin{figure}[!t]
		\centering
		\includegraphics[width=\columnwidth]{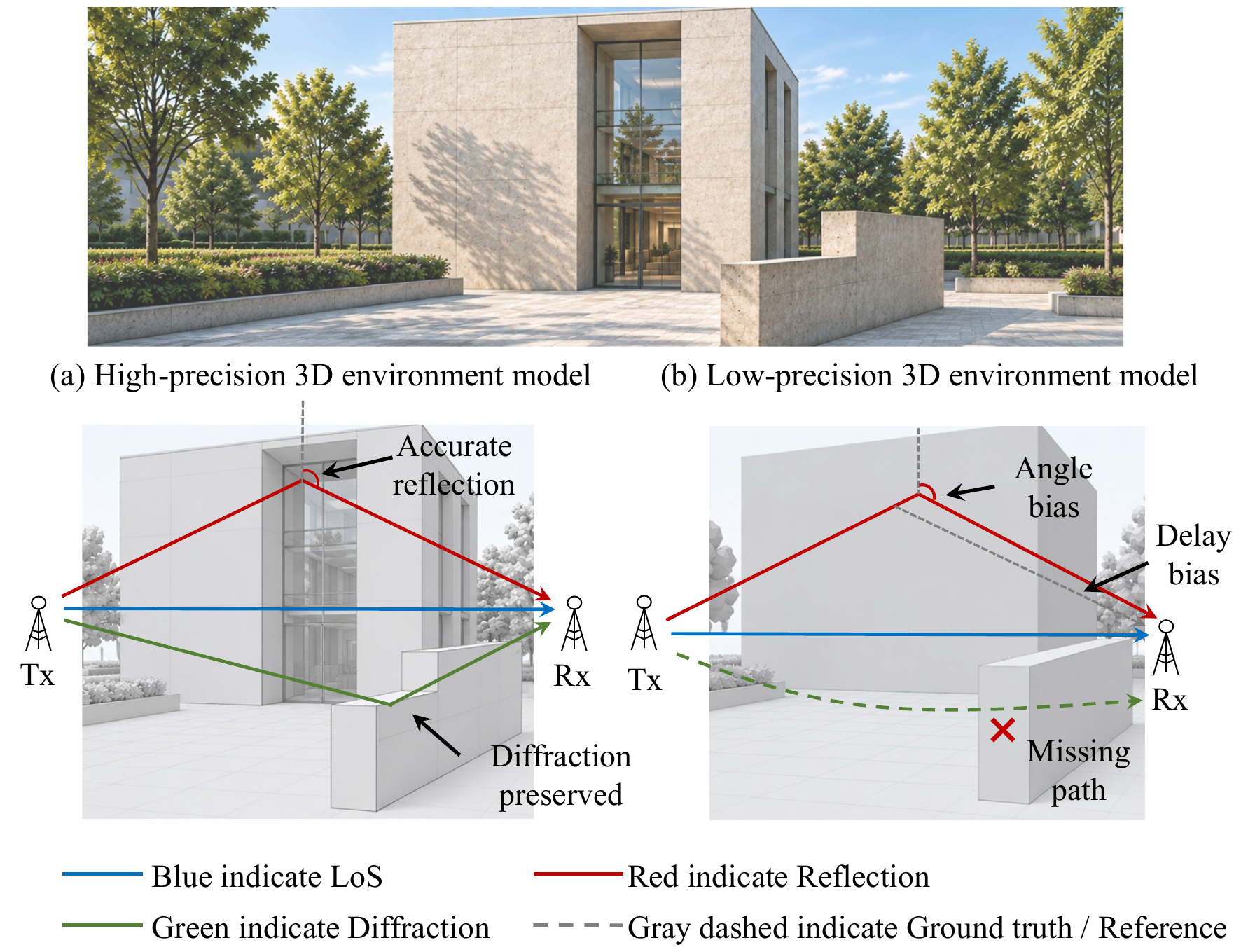}
		\caption{Conceptual comparison of multipath propagation in high- and low-precision three-dimensional environment models. (a) The high-precision model preserves reflection and diffraction paths. (b) The low-precision model introduces reflection-angle and delay biases and may miss a diffraction path, marked by the green dashed line and red cross. Blue, red, and green denote line-of-sight (LoS), reflection, and diffraction, respectively; the gray dashed line denotes the reference trajectory.}
		\label{fig:rt_channel_synthesis}
	\end{figure}

	\section{System Model and Problem Formulation}
	\label{sec:system_model}

	This section first introduces deterministic ray tracing (RT) and the synthesis of a multipath radio channel. The low-to-high-fidelity RT calibration problem is then formulated in terms of communication-relevant twin channel statistics, without assuming a one-to-one correspondence between individual multipath components.

	\begin{figure*}[!t]
		\centering
		\includegraphics[width=\textwidth]{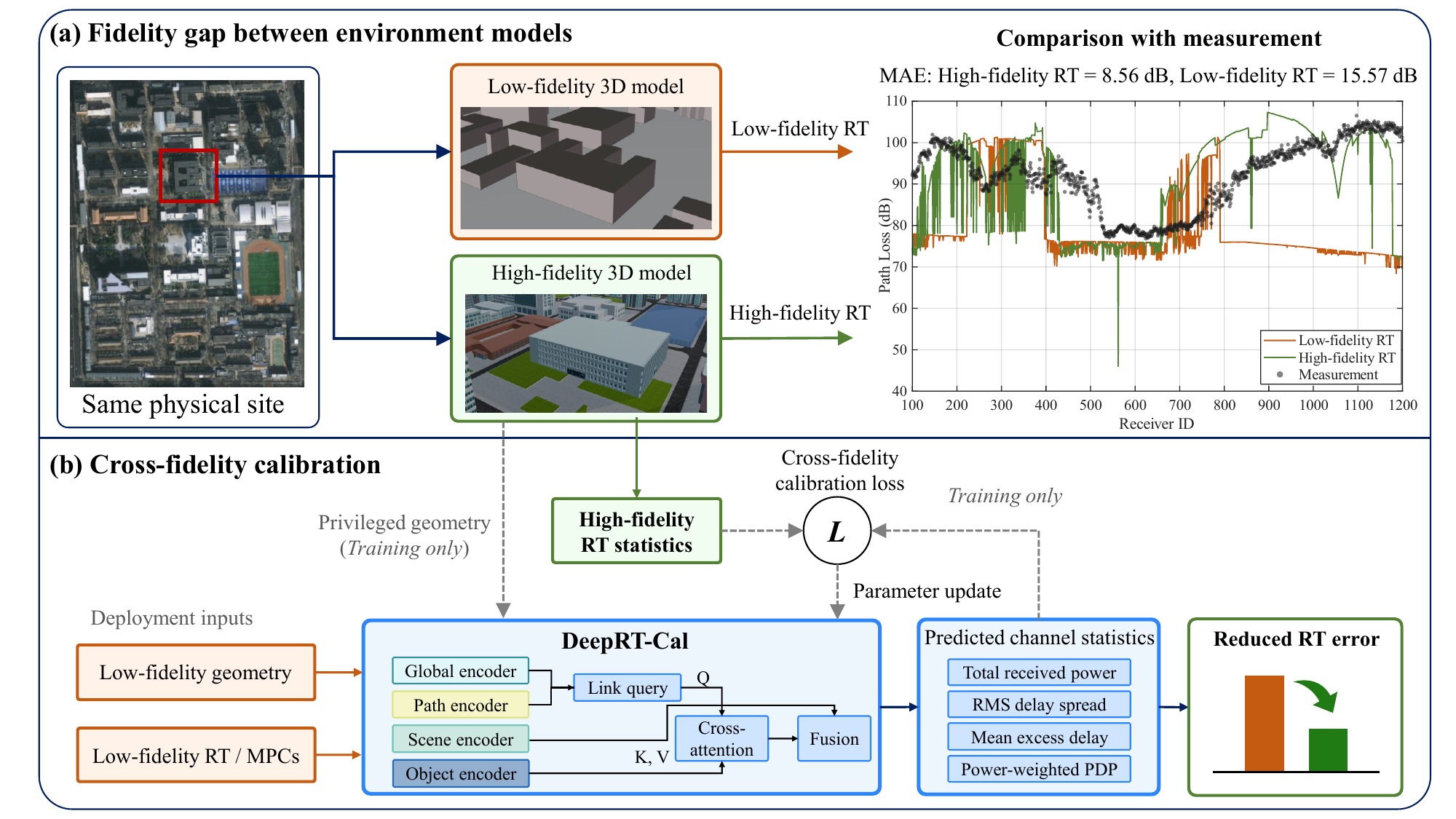}
		\caption{Overview of the cross-fidelity RT calibration problem and the proposed DeepRT-Cal framework. (a) High- and low-fidelity models of the same Beijing University of Posts and Telecommunications (BUPT) site produce uncalibrated RT results with different agreement to 1100 consecutive measurements. (b) High-fidelity geometry and RT statistics are used only during training, whereas deployment relies on low-fidelity geometry and RT/MPC features to predict high-fidelity-like channel statistics.}
		\label{fig:Architecture}
	\end{figure*}
	
	Fig.~\ref{fig:Architecture} provides a compact overview of the problem and the proposed solution. For the same BUPT campus region, the high-fidelity RT result follows the measurements more closely than the low-fidelity RT result over 1100 consecutive receiver locations; both curves are direct, uncalibrated RT outputs~\cite{Jiao2026WiWorldRealData}. DeepRT-Cal learns to compensate for this cross-fidelity gap by using low-fidelity geometry and RT/MPC features at deployment, while high-fidelity geometry and high-fidelity RT statistics are introduced only during training as privileged information and supervision, respectively. The network outputs high-fidelity-like channel statistics, including total received power, RMS delay spread, mean excess delay, and the power-weighted PDP.
	
	\subsection{Deterministic Ray Tracing and Multipath Channel Synthesis}
	\label{subsec:rt_channel_model}

	Ray tracing is a deterministic radio-channel modeling technique based on the high-frequency asymptotic approximation of electromagnetic wave propagation. Given a Tx, an Rx, a three-dimensional environment, and the associated electromagnetic and solver parameters, an RT solver searches for physically admissible propagation paths. Depending on the enabled propagation mechanisms, these paths may include line-of-sight (LoS) propagation, reflection, diffraction, penetration, and scattering \cite{scattering2, 14_scatter, diffraction2, UTD }. Fig.~\ref{fig:rt_channel_synthesis} illustrates how geometry changes the MPC set for the same Tx--Rx link. The high-precision scene in Fig.~\ref{fig:rt_channel_synthesis}(a) preserves reflection and diffraction paths. In Fig.~\ref{fig:rt_channel_synthesis}(b), simplified surfaces alter the reflection point and path length, causing angle and delay biases. Missing geometric details can also remove a diffraction path, as marked by the green dashed line and red cross. These changes affect both the multipath topology and the delay-power distribution.

	These profile-dependent MPC sets motivate the cross-fidelity formulation developed below.
	
	For the $s$th Tx--Rx link and RT configuration $q\in\{\mathrm{L},\mathrm{H}\}$, the detected multipath component (MPC) set is denoted by
	\begin{equation}
		\mathcal{P}_{s}^{q}
		=
		\left\{\mathbf{p}_{s,k}^{q}\right\}_{k=1}^{K_{s}^{q}},
	\end{equation}
	where $K_{s}^{q}$ is the number of detected MPCs. Each MPC is associated with a received power $P_{s,k}^{q}$, delay $\tau_{s,k}^{q}$, phase $\phi_{s,k}^{q}$, and propagation directions. If $L_{s,k}^{q}$ is the path length, its delay is $\tau_{s,k}^{q}=L_{s,k}^{q}/c$, where $c$ is the speed of light. The scalar channel impulse response (CIR) synthesized from these paths is
	\begin{equation}
		h_{s}^{q}(\tau)=
		\sum_{k=1}^{K_{s}^{q}}
		\sqrt{P_{s,k}^{q}}\exp\left(-\mathrm{j}\phi_{s,k}^{q}\right)
		\delta\left(\tau-\tau_{s,k}^{q}\right).
		\label{eq:cir}
	\end{equation}
	The corresponding power-delay profile (PDP) accumulates the received power over delay:
	\begin{equation}
		S_{s}^{q}(\tau)=
		\sum_{k=1}^{K_{s}^{q}}P_{s,k}^{q}\delta\left(\tau-\tau_{s,k}^{q}\right).
		\label{eq:pdp_continuous}
	\end{equation}

	For channel-level calibration, the total received power and the geometric excess delay are defined as
	\begin{align}
		P_{s,\mathrm{tot,dBm}}^{q}
		&=10\log_{10}\left(\frac{\sum_{k}P_{s,k}^{q}}{1~\mathrm{mW}}\right),\\
		\tau_{s,k}^{\mathrm{ex},q}
		&=\max\left(\tau_{s,k}^{q}-\tau_{s,0},0\right),
		\qquad
		\tau_{s,0}=\frac{\|\mathbf{r}_{s}-\mathbf{t}\|_{2}}{c}.
	\end{align}
	The absolute propagation delay is dominated by the Tx--Rx separation and is therefore not directly comparable across links with different distances. We subtract the geometric Tx--Rx propagation delay $\tau_{s,0}$ to isolate the additional delay introduced by reflection, diffraction, penetration, and other non-direct propagation mechanisms. The resulting excess-delay coordinate provides a common reference for comparing delay-domain structure across receiver locations. In particular, the first $10$-ns bin corresponds to paths whose excess length is less than approximately $3$~m and therefore indicates propagation close to the geometric Tx--Rx path. This reference is especially useful for LoS-dominated links, where the direct component can otherwise obscure differences among weaker multipath components.

	The power-weighted mean excess delay and RMS delay spread are respectively
	\begin{align}
		\bar{\tau}_{s,\mathrm{ex}}^{q}
		&=\frac{\sum_{k}P_{s,k}^{q}\tau_{s,k}^{\mathrm{ex},q}}
		{\sum_{k}P_{s,k}^{q}},\\
		\sigma_{\tau,s}^{q}
		&=\sqrt{\frac{\sum_{k}P_{s,k}^{q}
				\left(\tau_{s,k}^{\mathrm{ex},q}-\bar{\tau}_{s,\mathrm{ex}}^{q}\right)^{2}}
			{\sum_{k}P_{s,k}^{q}}}.
	\end{align}
	
	For the distributional target, the excess-delay axis is restricted to the evaluation window $[0~\mathrm{ns},2000~\mathrm{ns})$ and partitioned into $N_{\tau}=200$ bins of width $\Delta\tau=10$~ns, where $\mathcal{B}_{b}=[(b-1)\Delta\tau,b\Delta\tau)$ and $b=1,\ldots,N_{\tau}$. Let
	\(\mathcal{K}_{s,W}^{q}=\{k:\tau_{s,k}^{\mathrm{ex},q}\in[0~\mathrm{ns},2000~\mathrm{ns})\}\)
	denote the paths inside this window. Their powers are accumulated within the delay bins and normalized to form the power-weighted PDP; paths outside the stated evaluation window are excluded before normalization. The unique numbered definition used for evaluation is given in Section~\ref{subsec:results_discussion}.
	
	\subsection{Cross-Fidelity RT Calibration Problem}
	\label{subsec:problem_formulation}
	
	Consider two RT configurations for the same Tx--Rx link. The low-fidelity configuration, denoted by $q=\mathrm{L}$, uses a simplified environmental representation and is computationally efficient. The high-fidelity configuration, denoted by $q=\mathrm{H}$, uses a more detailed representation and serves as the reference channel. The two configurations share the Tx--Rx locations and the relevant radio settings, but their geometric descriptions may produce different propagation topologies and MPC sets.
	
	For the $s$th link, the low- and high-fidelity MPC sets are written as $\mathcal{P}_{s}^{\mathrm{L}}$ and $\mathcal{P}_{s}^{\mathrm{H}}$, respectively. In general,
	\begin{equation}
		K_{s}^{\mathrm{L}}\neq K_{s}^{\mathrm{H}},
	\end{equation}
	and an individual low-fidelity MPC does not necessarily have a unique high-fidelity counterpart. Geometric simplification may remove valid paths, introduce additional paths, or substantially alter the delay and power of the remaining paths. Therefore, directly learning a path-by-path mapping from $\mathcal{P}_{s}^{\mathrm{L}}$ to $\mathcal{P}_{s}^{\mathrm{H}}$ is ill posed and requires an unstable correspondence assumption.
	
	
	The deployment input for the $s$th link is
	\begin{equation}
		\mathbf{x}_{s}^{\mathrm{D}}
		=\left(
		\mathbf{g}_{s},
		\mathbf{X}_{s}^{\mathrm{P,L}},
		\mathbf{X}_{s}^{\mathrm{S,L}},
		\mathbf{X}_{s}^{\mathrm{O,L}}
		\right),
		\label{eq:deployment_input}
	\end{equation}
	where $\mathbf{g}_{s}$ is the global low-fidelity link descriptor and the superscript $\mathrm{L}$ identifies low-fidelity geometry or RT information. During training, high-fidelity geometric descriptors are available as privileged information. A structured group mask $\mathbf{m}_{s}$ is applied to the high-fidelity scene and object groups, giving
	\begin{equation}
		\widetilde{\mathbf{x}}_{s}^{\mathrm{tr}}
		=\mathcal{M}\!\left(
		\mathbf{x}_{s}^{\mathrm{D}},
		\mathbf{X}_{s}^{\mathrm{S,H}},
		\mathbf{X}_{s}^{\mathrm{O,H}};\mathbf{m}_{s}
		\right).
		\label{eq:privileged_training_input}
	\end{equation}
	The high-fidelity groups are masked or dropped during training so that the network does not depend on their presence. At inference, the same groups are absent or set to their masked values, and the network receives only $\mathbf{x}_{s}^{\mathrm{D}}$.
	
	The fixed-dimensional high-fidelity supervision target is
	\begin{equation}
		\mathbf{y}_{s}^{\mathrm{H}}
		=\left[
		P_{s,\mathrm{tot,dBm}}^{\mathrm{H}},
		\;\sigma_{\tau,s}^{\mathrm{H}},
		\;\bar{\tau}_{s,\mathrm{ex}}^{\mathrm{H}},
		\;\left(\mathbf{p}_{s}^{\mathrm{H}}\right)^{\mathsf{T}}
		\right]^{\mathsf{T}},
		\label{eq:high_fidelity_target}
	\end{equation}
	where $\mathbf{p}_{s}^{\mathrm{H}}\in\mathbb{R}^{N_{\tau}}$ is the normalized high-fidelity power-weighted PDP. DeepRT-Cal learns
	\begin{equation}
		\widehat{\mathbf{y}}_{s}^{\mathrm{H}}
		=f_{\theta}\!\left(\widetilde{\mathbf{x}}_{s}^{\mathrm{tr}}\right),
		\label{eq:calibration_mapping}
	\end{equation}
	where the network implementation separates the first three scalar components from the $N_{\tau}$-dimensional PDP component through two prediction heads. The mapping is evaluated at deployment with the high-fidelity groups masked.
	
	For $N$ paired training samples, the training set and optimization problem are written as
	\begin{align}
		\mathcal{D}
		&=\left\{\left(\widetilde{\mathbf{x}}_{s}^{\mathrm{tr}},\mathbf{y}_{s}^{\mathrm{H}}\right)\right\}_{s=1}^{N},\\
		\theta^{*}
		&=\arg\min_{\theta}\frac{1}{N}\sum_{s=1}^{N}
		\mathcal{L}\!\left(
		f_{\theta}\!\left(\widetilde{\mathbf{x}}_{s}^{\mathrm{tr}}\right),
		\mathbf{y}_{s}^{\mathrm{H}}
		\right).
		\label{eq:calibration_objective}
	\end{align}
	
	At deployment, the same learned mapping is evaluated with the high-fidelity groups masked, so that the effective input is $\mathbf{x}_{s}^{\mathrm{D}}$.
	
\begin{figure*}[!t]
	\centering
	
	\subfloat[Global, path, and scene feature construction.]
	{
		\label{fig:feature}
		\includegraphics[width=0.7\textwidth]{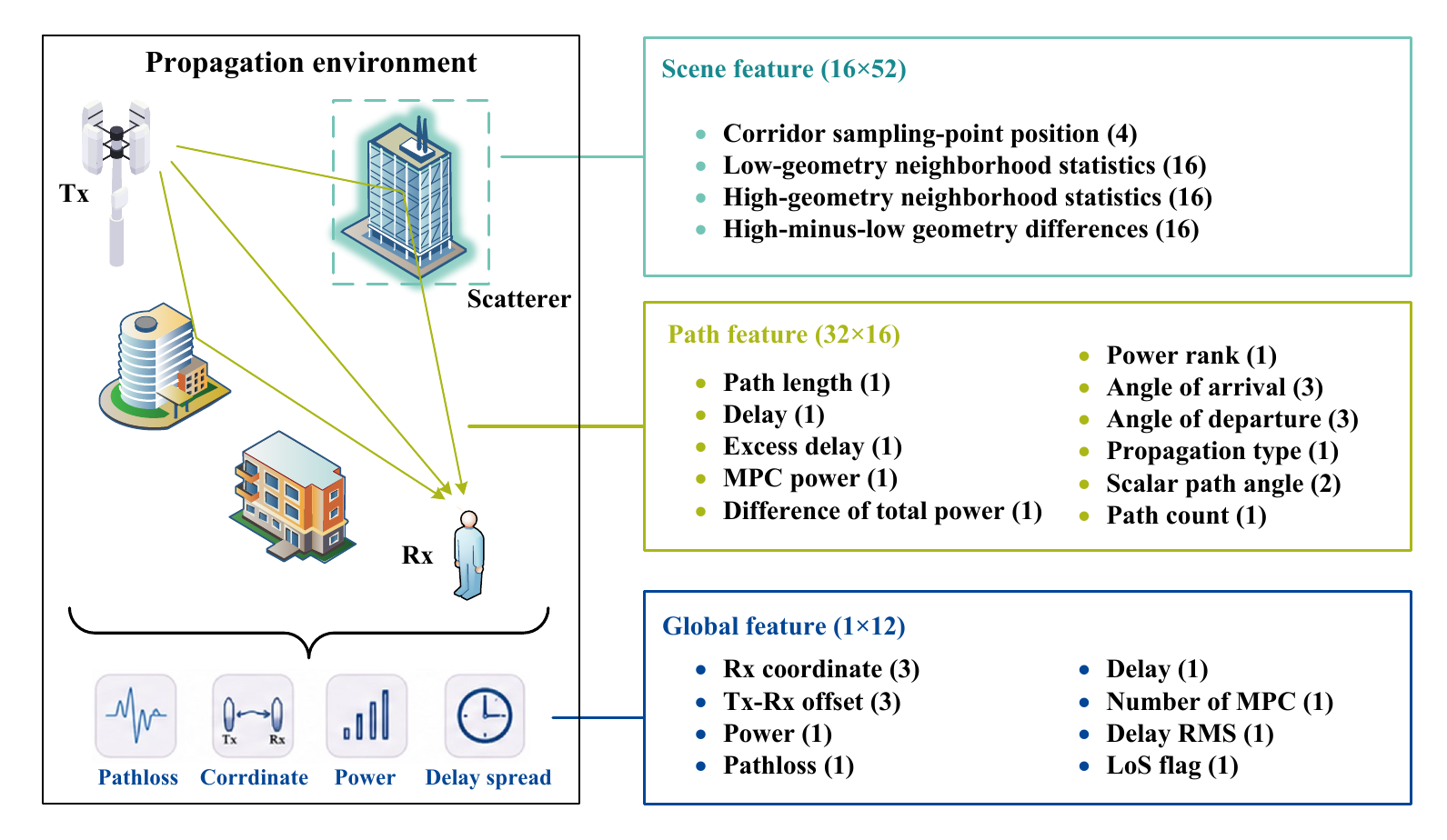}
	}
	
	\vspace{1.5mm}
	
	\subfloat[Link-centered object-token construction.]
	{
		\label{fig:object_token}
		\includegraphics[width=0.7\textwidth]{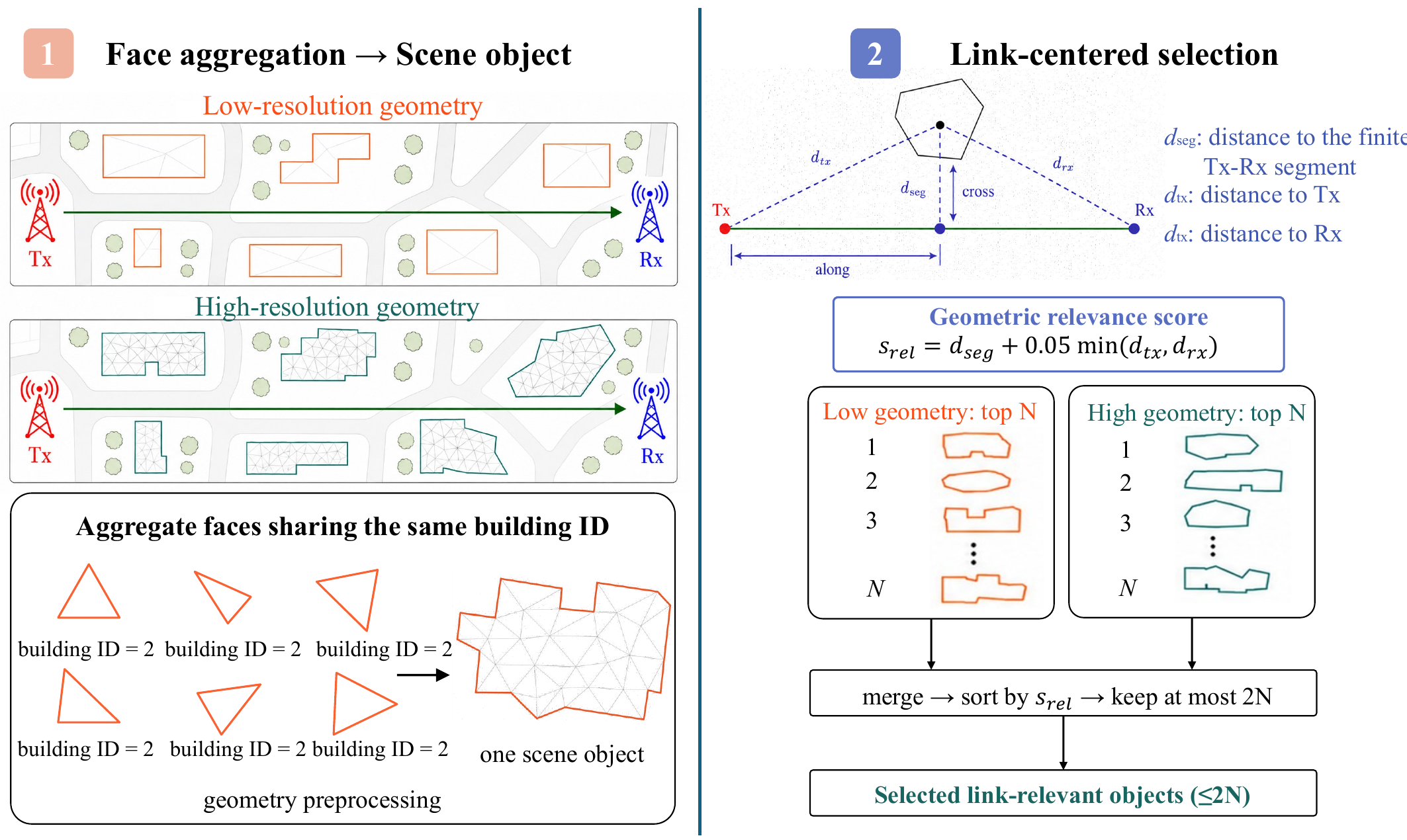}
	}
	
	\caption{Heterogeneous feature construction for one Tx--Rx link.
		(a) Global link features $(1\times12)$, low-fidelity path features $(32\times16)$, and propagation-corridor scene features $(16\times52)$.
		(b) Link-centered object construction: faces sharing an object ID are aggregated, ranked by geometric relevance, and retained as at most 64 tokens with 23 features each.
		High-geometry features are privileged training inputs and are masked or removed at deployment.}
	\label{fig:feature_object}
\end{figure*}

	
\subsection{Heterogeneous Feature Representation}
\label{subsec:feature_representation}

The deployment input in \eqref{eq:deployment_input} is heterogeneous in both physical meaning and cardinality. As illustrated in Fig.~\ref{fig:feature_object}, the information available for one Tx--Rx link is organized into four groups: a global link descriptor, a set of low-fidelity MPC tokens, propagation-corridor scene tokens, and link-centered environmental object tokens. These groups differ in physical meaning, feature dimension, and token cardinality. The MPC and object groups also have set structure, so their elements should not be interpreted as fixed-coordinate entries or assumed to have one-to-one correspondence across fidelity profiles. Directly concatenating the raw groups would therefore obscure their distinct roles and would make the representation sensitive to arbitrary padding or ordering.

The global feature group describes the macroscopic state of the link, including the Rx coordinate, Tx--Rx offset and distance-related quantities, geometric delay, low-fidelity power and delay spread, MPC count, and an LoS indicator. It provides a stable link-level context for conditioning the later query and fusion operations. The path feature group records the low-fidelity RT evidence carried by the strongest retained MPCs. Each token contains path geometry, delay, excess delay, power, propagation type, direction information, and rank-related features. The paths are padded to a fixed maximum cardinality for batch processing, while the padding mask identifies valid tokens. Sorting is used only to retain dominant paths and construct the rank feature; it does not establish a correspondence with any high-fidelity path.

The scene feature group samples local geometry along the Tx--Rx corridor. At each sampling position, the representation contains the sampling location, low-geometry neighborhood statistics, high-geometry neighborhood statistics, and their high-minus-low differences. The high-geometry terms expose the geometric discrepancy during training, but they are subject to the structured masking operation in \eqref{eq:privileged_training_input} and are absent at deployment. This design allows the network to learn from detailed geometry during optimization without making detailed geometry a required inference input.

The object branch requires a link-centered selection because a complete scene mesh contains many structures unrelated to a particular link. As shown in Fig.~\ref{fig:feature_object}(b), triangular faces sharing a building or object identifier are first aggregated into scene objects. Each object is then assigned the geometric relevance score
\begin{equation}
	s_{\mathrm{rel}}
	=
	d_{\mathrm{seg}}+0.05\min(d_{\mathrm{tx}},d_{\mathrm{rx}}),
	\label{eq:object_relevance}
\end{equation}
where $d_{\mathrm{seg}}$ is the distance from the object center to the finite Tx--Rx segment, and $d_{\mathrm{tx}}$ and $d_{\mathrm{rx}}$ are the distances to its endpoints. Objects are ranked independently for the two geometry profiles; the top $N$ from each profile are merged and truncated to at most $2N$ link-relevant objects. The high-fidelity candidates are governed by the structured group mask during training and are unavailable at inference.

Fig.~\ref{fig:feature_object}(a) summarizes the global, path, and scene features, while Fig.~\ref{fig:feature_object}(b) details object selection and tokenization. The next section describes how the four groups are encoded and fused for channel-statistics prediction.
	
\section{Proposed Link-Query Cross-Attention Network}
\label{sec:proposed_network}

\subsection{Architecture Overview}

\begin{figure*}[!t]
	\centering
	\includegraphics[width=\textwidth]{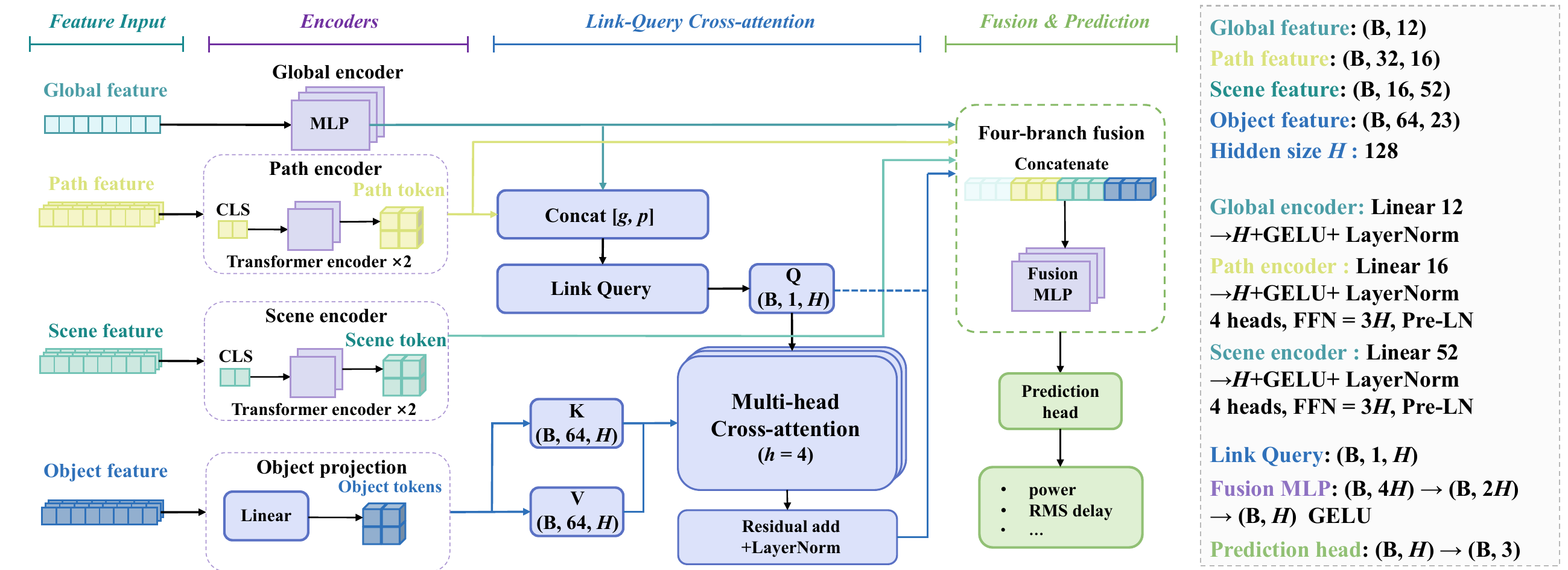}
	\caption{Architecture of the proposed link-query cross-attention network. Global, low-fidelity MPC, and scene-corridor features are independently encoded. A link query constructed from the global and MPC-set representations attends to the link-centered object tokens. The resulting object context is fused with the other three branches to predict high-fidelity twin channel statistics.}
	\label{fig:net_arc}
\end{figure*}

The calibration task considered in this work involves four heterogeneous sources of information: link-level descriptors, a variable-cardinality set of low-fidelity multipath components (MPCs), geometric samples along the transmitter--receiver (Tx--Rx) corridor, and nearby environmental objects. These inputs differ in physical meaning, tensor organization, and cardinality. Directly concatenating their raw features would ignore the permutation structure of the MPC and object sets and would also make it difficult to determine which environmental objects are relevant to a particular link. We therefore propose the link-query cross-attention network illustrated in Fig.~\ref{fig:net_arc}.

As shown from left to right in Fig.~\ref{fig:net_arc}, the network contains four stages. First, the feature-input stage constructs global link features, low-fidelity MPC tokens, propagation-corridor scene tokens, and link-centered object tokens. During training, high-fidelity geometric groups may be supplied as privileged features but are subject to structured masking. Second, independent encoders map the heterogeneous inputs to a common $H$-dimensional latent space. Third, a link query formed from the global and low-fidelity MPC representations retrieves link-relevant object context through multi-head cross-attention. Finally, the global, path, scene, and attended-object representations are fused to predict three scalar high-fidelity statistics and a normalized power-weighted excess-delay distribution.

For a mini-batch of size $B$, the four input tensors are denoted by
\begin{align}
	\mathbf{G}&\in\mathbb{R}^{B\times 12},
	&\mathbf{X}^{P}&\in\mathbb{R}^{B\times K_{\max}\times16},\nonumber\\
	\mathbf{X}^{S}&\in\mathbb{R}^{B\times N_S\times52},
	&\mathbf{X}^{O}&\in\mathbb{R}^{B\times O_{\max}\times23},
	\label{eq:network_inputs}
\end{align}
where $K_{\max}=32$, $N_S=16$, and $O_{\max}=64$. Boolean masks $\mathbf{M}^{P}$ and $\mathbf{M}^{O}$ distinguish valid path and object tokens from zero padding. All continuous features are standardized using means and standard deviations estimated exclusively from the training split. The network shown in Fig.~\ref{fig:net_arc} uses hidden dimension $H=128$, four attention heads, and two Transformer layers in each set encoder.

\subsection{Feature Encoders}

Throughout this section, $\mathrm{LN}$, $\mathrm{GELU}$, and $\mathrm{MLP}$ denote layer normalization, the Gaussian error linear unit, and a multilayer perceptron, respectively. Upright roman type distinguishes these operations from variables. Features and embeddings are row vectors, tokens are stacked as rows, and $\mathbf{W}$ and $\mathbf{b}$ denote learnable weights and biases.

\subsubsection{Global Encoder}

The global branch describes the macroscopic state of one Tx--Rx link. Its 12-dimensional input is
\begin{align}
	\mathbf{g}_s=[&\mathbf{r}_s,\ \mathbf{r}_s-\mathbf{t},\ d_s,\ \tau_{s,0},\nonumber\\
	&P_{s,\mathrm{tot,dBm}}^{\mathrm{L}},\ \log\!\left(1+\frac{\sigma_{\tau,s}^{\mathrm{L}}}{1~\mathrm{ns}}\right),\nonumber\\
	&\log(1+K_s^{\mathrm{L}}),\ I_{\mathrm{LoS},s}^{\mathrm{L}}],
	\label{eq:global_input_ch3}
\end{align}
where $\mathbf{t}$ and $\mathbf{r}_s$ are the Tx and Rx coordinates, respectively, $d_s=\lVert\mathbf{r}_s-\mathbf{t}\rVert_2$, $\tau_{s,0}=d_s/c$ is the geometric Tx--Rx propagation delay, $P_{s,\mathrm{tot,dBm}}^{\mathrm{L}}$ is the low-fidelity total received power in dBm, $\sigma_{\tau,s}^{\mathrm{L}}$ is the low-fidelity root-mean-square (RMS) delay spread, $K_s^{\mathrm{L}}$ is the low-fidelity path count, and $I_{\mathrm{LoS},s}^{\mathrm{L}}$ is an LoS indicator.

After feature-wise normalization, the global vector is projected to the common latent dimension:
\begin{equation}
	\mathbf{z}_s^{G}=\mathrm{LN}\left(
	\mathrm{GELU}\left(\bar{\mathbf{g}}_s\mathbf{W}_{G}+\mathbf{b}_{G}\right)
	\right)\in\mathbb{R}^{1\times H},
	\label{eq:global_encoder_ch3}
\end{equation}
where the overbar denotes standardized features. This branch provides a stable link-level reference for both query construction and final feature fusion.

\subsubsection{Path Encoder}

The path branch retains at most the $K_{\max}=32$ strongest low-fidelity MPCs, sorted by linear received power before padding. Each 16-dimensional path token contains log path length, log propagation delay, log excess delay, received power in dBm, power relative to the low-fidelity total, propagation type, sine and cosine of the scalar path angle, three-dimensional departure and arrival directions, normalized power rank, and log total low-fidelity path count. Length and delay features use $\log(1+x/x_0)$ with $x_0=1$~m and $1$~ns, respectively. Sorting is used only to select the dominant paths and construct the rank feature; it does not establish a correspondence with any high-fidelity path.

Each standardized token is first projected to $H$ dimensions. A learnable classification token $\mathbf{c}^{P}$ is then prepended:
\begin{equation}
	\mathbf{H}_{s}^{P,0}=\begin{bmatrix}
	\mathbf{c}^{P}\\
	\mathrm{LN}\left(
	\mathrm{GELU}\left(\bar{\mathbf{X}}_{s}^{P}\mathbf{W}_{P}+\mathbf{b}_{P}\right)
	\right)
	\end{bmatrix}.
	\label{eq:path_projection_ch3}
\end{equation}
The sequence is processed by two pre-normalized Transformer encoder layers~\cite{Vaswani2017Attention}. Each layer uses standard four-head self-attention with padding mask $\mathbf{A}^{P}$, per-head dimension $d_h=H/4$, and a feed-forward dimension of $3H$. The additive mask $\mathbf{A}^{P}$ is derived from $\mathbf{M}^{P}$: valid keys receive 0 and padding keys receive $-\infty$, with the CLS token always valid. The normalized CLS output forms the path-set representation
\begin{equation}
	\mathbf{z}_s^{P}=\mathrm{LN}\left(
	\mathbf{H}_{s}^{P,2}[0]
	\right)\in\mathbb{R}^{1\times H}.
	\label{eq:path_representation_ch3}
\end{equation}
This operation summarizes the complete retained MPC set without requiring low-/high-fidelity path matching.

\subsubsection{Scene Encoder}

The scene branch describes geometric variation along the direct Tx--Rx corridor. Sixteen locations are sampled along the link. At each location, local vertex neighborhoods are summarized separately in the simplified and detailed geometries. The resulting 52-dimensional scene token contains the normalized sampling position and its three-dimensional coordinates, 16 simplified-geometry descriptors, 16 detailed-geometry descriptors, and their 16-dimensional difference:
\begin{equation}
	\mathbf{x}_{s,i}^{S}=\left[
	t_i,x_i,y_i,z_i,
	\mathbf{u}_{s,i}^{\mathrm{L}},
	\mathbf{u}_{s,i}^{\mathrm{H}},
	\mathbf{u}_{s,i}^{\mathrm{H}}-\mathbf{u}_{s,i}^{\mathrm{L}}
	\right]\in\mathbb{R}^{1\times52}.
	\label{eq:scene_token_ch3}
\end{equation}
Here, $i=1,\ldots,N_S$ indexes corridor samples, $t_i\in[0,1]$ is the normalized link position, and $(x_i,y_i,z_i)$ are the sample coordinates. The local descriptors summarize horizontal distance, absolute and link-relative height, and obstruction statistics around the sampled location. The explicit difference term exposes geometry changes that may alter reflection, diffraction, or blockage conditions, but it does not use any high-fidelity MPC label.

As illustrated by the scene encoder in Fig.~\ref{fig:net_arc}, the 16 tokens are projected to $H$ dimensions, prepended with a learnable classification token $\mathbf{c}^{S}$, and processed by two four-head pre-normalized Transformer layers with feed-forward dimension $3H$. The final classification-token output is
\begin{equation}
	\mathbf{z}_s^{S}=\mathrm{LN}\left(
	\mathbf{H}_{s}^{S,2}[0]
	\right)\in\mathbb{R}^{1\times H}.
	\label{eq:scene_representation_ch3}
\end{equation}
Because the normalized link position $t_i$ is included in each token, the encoder can distinguish geometric conditions near the Tx, inside the propagation corridor, and near the Rx.

\subsubsection{Object Projection}

The object branch provides a compact representation of nearby buildings and wall structures. Triangular faces belonging to the same environmental object are aggregated before per-link tokenization. Each 23-dimensional token records the geometry profile, normalized along-link and cross-link positions, distances to the Tx, Rx, and Tx--Rx segment, vertical position and extent, horizontal bounding-box dimensions, surface area, face count, average normal, material descriptor, scattering fraction, and link-proximity indicators. Up to $N$ objects are retained from each geometry, yielding at most $O_{\max}=64$ tokens for one link.

Unlike the path and scene branches, the object branch does not perform self-attention before interacting with the link query. The standardized object features are independently projected into the common latent space:
\begin{equation}
	\mathbf{E}_s=\mathrm{LN}\left(
	\mathrm{GELU}\left(
	\bar{\mathbf{X}}_s^{O}\mathbf{W}_{E}+\mathbf{b}_{E}
	\right)\right)
	\in\mathbb{R}^{O_{\max}\times H}.
	\label{eq:object_projection_ch3}
\end{equation}
The resulting matrix $\mathbf{E}_s$ supplies the keys and values in the cross-attention stage. Retaining individual object tokens, rather than prematurely averaging them, allows the query to assign different relevance to different environmental structures.

\subsection{Link-Query Cross-Attention}

The central module in Fig.~\ref{fig:net_arc} is designed to answer a link-conditioned question: given the current Tx--Rx configuration and the low-fidelity MPC evidence, which nearby environmental objects are most relevant to the desired channel correction? A single query is constructed by concatenating the global and path-set embeddings and applying a query MLP:
\begin{equation}
	\mathbf{q}_s=\mathrm{MLP}_{Q}\left(
	[\mathbf{z}_s^{G},\mathbf{z}_s^{P}]
	\right)\in\mathbb{R}^{1\times H}.
	\label{eq:link_query_ch3}
\end{equation}
The embeddings are concatenated along the feature dimension. For a batch, the query tensor has shape $(B,1,H)$. The scene representation is intentionally not used to construct the query; it remains an independent branch for final fusion. This separation lets the object-attention module be driven directly by the link state and observed low-fidelity propagation.

For attention head $m\in\{1,\ldots,4\}$, the projected query, keys, and values are
\begin{align}
	\mathbf{Q}_{s,m}&=\mathbf{q}_s\mathbf{W}_{Q,m},
	&\mathbf{K}_{s,m}&=\mathbf{E}_s\mathbf{W}_{K,m},\nonumber\\
	\mathbf{V}_{s,m}&=\mathbf{E}_s\mathbf{W}_{V,m},
	\label{eq:qkv_ch3}
\end{align}
where $\mathbf{Q}_{s,m}\in\mathbb{R}^{1\times d_h}$ and $\mathbf{K}_{s,m},\mathbf{V}_{s,m}\in\mathbb{R}^{O_{\max}\times d_h}$. The scaled dot-product attention output~\cite{Vaswani2017Attention} is
\begin{equation}
	\mathbf{C}_{s,m}=\mathrm{softmax}\left(
	\frac{\mathbf{Q}_{s,m}\mathbf{K}_{s,m}^{\mathsf{T}}}{\sqrt{d_h}}
	+\mathbf{A}_s^{O}
	\right)\mathbf{V}_{s,m},
	\label{eq:cross_attention_ch3}
\end{equation}
where $\mathbf{A}_s^{O}\in\{0,-\infty\}^{1\times O_{\max}}$ is derived from $\mathbf{M}_s^{O}$, with 0 for valid objects and $-\infty$ for padding, and is shared across heads. This attention operation requires at least one valid object. Outputs from the four heads are concatenated and linearly projected:
\begin{equation}
	\mathbf{C}_s=\mathrm{Concat}
	(\mathbf{C}_{s,1},\ldots,\mathbf{C}_{s,4})\mathbf{W}_{C}.
	\label{eq:multihead_context_ch3}
\end{equation}
Following the residual path shown below the cross-attention block in Fig.~\ref{fig:net_arc}, the final object-context representation is
\begin{equation}
	\mathbf{z}_s^{O}=\mathrm{LN}
	(\mathbf{q}_s+\mathbf{C}_s)\in\mathbb{R}^{1\times H}.
	\label{eq:object_context_ch3}
\end{equation}

The module produces one object-context vector for the complete link, rather than one output for every MPC. The attention weights quantify learned relevance within the statistical prediction task; they are not intended as path-level attribution to physical scatterers.

\subsection{Four-Branch Fusion and Channel-Statistics Prediction}

The right-hand side of Fig.~\ref{fig:net_arc} combines the four complementary link representations. The global branch describes the macroscopic link state, the path branch summarizes the low-fidelity MPC set, the scene branch represents geometric changes along the propagation corridor, and the object branch provides query-conditioned environmental context. Their concatenation is
\begin{equation}
	\mathbf{u}_s=[\mathbf{z}_s^{G},\mathbf{z}_s^{P},
	\mathbf{z}_s^{S},\mathbf{z}_s^{O}]
	\in\mathbb{R}^{1\times4H}.
	\label{eq:four_branch_concat_ch3}
\end{equation}
The fusion MLP follows the dimensional sequence $4H\rightarrow2H\rightarrow H$:
\begin{align}
	\mathbf{h}_s^{F}&=\mathrm{LN}\left(
	\mathrm{GELU}\left(\mathbf{u}_s\mathbf{W}_{F1}+\mathbf{b}_{F1}\right)
	\right),\nonumber\\
	\mathbf{z}_s^{F}&=\mathrm{LN}\left(
	\mathrm{GELU}\left(\mathbf{h}_s^{F}\mathbf{W}_{F2}+\mathbf{b}_{F2}\right)
	\right)\in\mathbb{R}^{1\times H}.
	\label{eq:fusion_mlp_ch3}
\end{align}
This late-fusion strategy preserves the role of each physical information source until a link-level representation has been formed.

Define the common dimensionless transform for the two positive delay targets as
\begin{equation}
	T(x)=\log\!\left(1+\frac{x}{1~\mathrm{ns}}\right).
	\label{eq:positive_delay_transforms}
\end{equation}

The scalar statistics head is a linear mapping from the fused representation to three standardized targets:
\begin{equation}
	\widehat{\widetilde{\mathbf{y}}}_{s}^{\,\mathrm{sc}}
	=\mathbf{z}_{s}^{F}\mathbf{W}_{Y}+\mathbf{b}_{Y}
	\in\mathbb{R}^{1\times3},
	\label{eq:prediction_head_ch3}
\end{equation}
where the corresponding physical target vector is
\begin{equation}
	\mathbf{y}_{s}^{\mathrm{sc}}
	=\left[
	P_{s,\mathrm{tot,dBm}}^{\mathrm{H}},
	\;T\!\left(\sigma_{\tau,s}^{\mathrm{H}}\right),
	\;T\!\left(\bar{\tau}_{s,\mathrm{ex}}^{\mathrm{H}}\right)
	\right].
	\label{eq:prediction_targets_ch3}
\end{equation}

The auxiliary power-delay head produces a probability vector over the $N_{\tau}=200$ excess-delay bins:
\begin{equation}
	\widehat{\mathbf{p}}_{s}^{\mathrm{H}}
	=\operatorname{softmax}\!\left(
	\mathbf{z}_{s}^{F}\mathbf{W}_{\mathrm{PDP}}+\mathbf{b}_{\mathrm{PDP}}
	\right)
	\in\Delta^{N_{\tau}-1},
	\label{eq:pdp_prediction_head}
\end{equation}
where $\Delta^{N_{\tau}-1}$ denotes the probability simplex. The power-weighted PDP is therefore reconstructed from the distribution head rather than treated as a collection of independently regressed scalar outputs. At inference, standardization and the $\log(1+x)$ transform are inverted, and the predicted delay statistics are clipped to nonnegative values.

\subsection{Loss Function and Training Strategy}

Let $\widetilde{y}_{s,j}$ and $\widehat{\widetilde{y}}_{s,j}$ be the standardized target and prediction for scalar task $j\in\{P,\sigma,\mu\}$. The scalar regression losses $\mathcal{L}_{j}$ use batch-averaged Smooth-$L_{1}$ penalties. The PDP loss $\mathcal{L}_{\mathrm{PDP}}$ is the batch-averaged Kullback--Leibler divergence from the high-fidelity target distribution to the predicted distribution, with a small positive constant $\varepsilon$ for numerical stability. The complete objective is
\begin{equation}
	\mathcal{L}_{\mathrm{total}}
	=\lambda_{P}\mathcal{L}_{P}
	+\lambda_{\sigma}\mathcal{L}_{\sigma}
	+\lambda_{\mu}\mathcal{L}_{\mu}
	+\lambda_{\mathrm{PDP}}\mathcal{L}_{\mathrm{PDP}}.
	\label{eq:total_loss_ch3}
\end{equation}

The four coefficients are specified in the simulation setup. Feature normalization statistics and scalar-target normalization statistics are estimated from the training data only and are fixed for validation and testing. The path and object masks prevent padded or unavailable groups from contributing to attention.

	\section{Simulation Results and Discussion}
	
	\subsection{Simulation Setup and Parameter Configuration}
	\label{subsec:simulation_setup}
	
	\subsubsection{Paired Campus Environments}
	
	\begin{figure}[!t]
		\centering
		\includegraphics[width=\columnwidth]{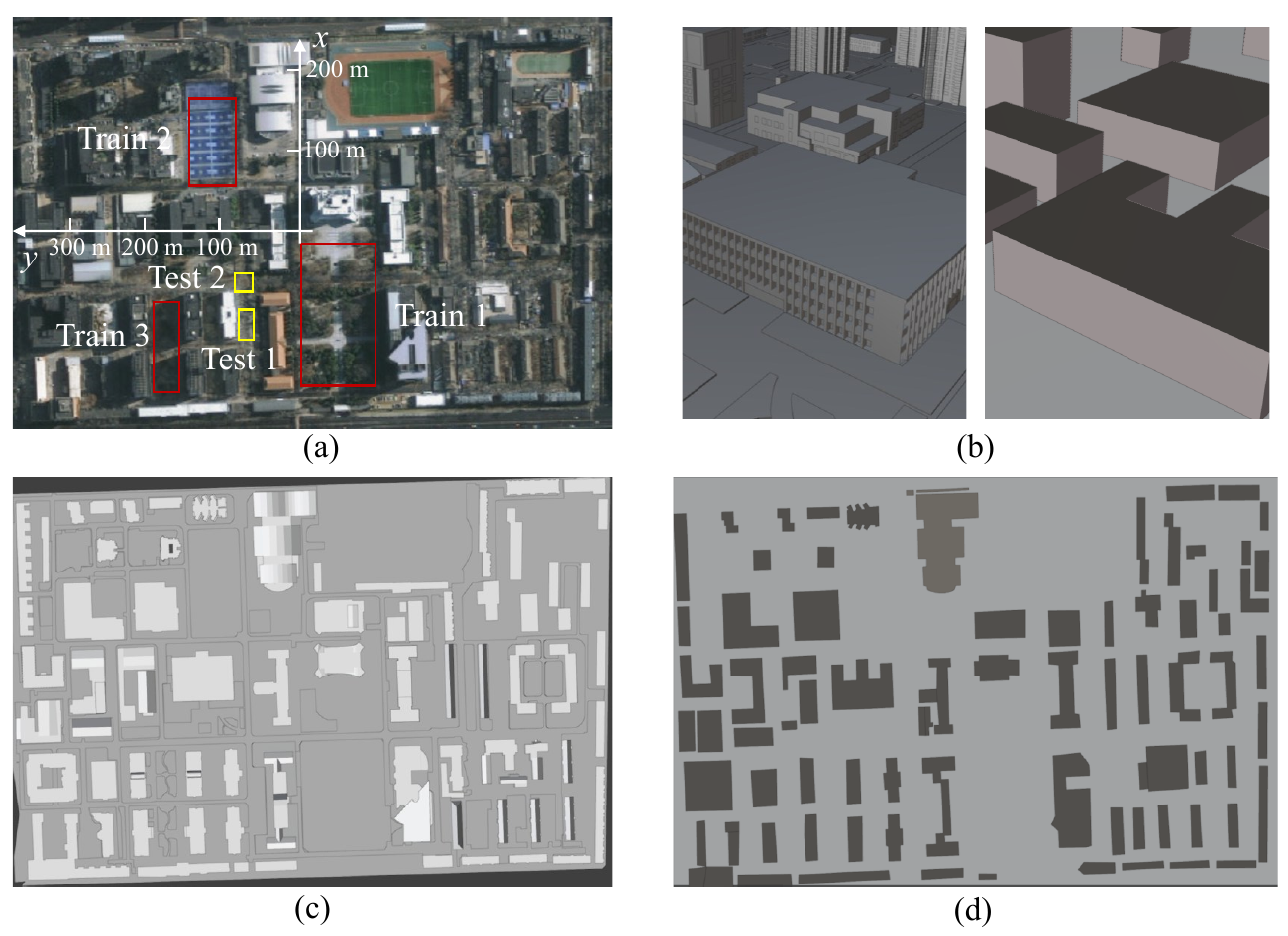}
		\caption{BUPT campus environment and paired scene models used for RT simulation. (a) Satellite view of the campus and the five receiver regions. (b) Co-located local comparison between the high-fidelity geometry on the left and the low-fidelity geometry on the right. (c) Top view of the complete high-fidelity scene. (d) Top view of the complete low-fidelity OSM-based scene. The same Cartesian coordinate system, Tx location, and receiver-region annotations are used in all panels.}
		\label{fig:bupt_scenes}
	\end{figure}
	
	The simulations are conducted in an outdoor campus environment at BUPT. Two geometrically aligned scene models are prepared for each Tx--Rx link: a low-fidelity scene and a high-fidelity scene. The low-fidelity scene is generated from OSM data and preserves the dominant building footprints and coarse height information \cite{openstreetmap}. It represents the type of openly accessible map that can be obtained at relatively low cost without dedicated site surveying. The high-fidelity scene is reconstructed from simultaneous localization and mapping (SLAM) and Global Navigation Satellite System (GNSS)-assisted measurements. After post-processing, its geometric error is controlled below $10$~cm, allowing substantially finer building boundaries, facade structures, and roof details to be retained.
	
	The two scene models are registered to the same Cartesian coordinate system. Consequently, an identical Tx--Rx pair can be evaluated in both RT profiles without an additional geometric registration step. Fig.~\ref{fig:bupt_scenes} summarizes the simulated environment and the paired scene representations. Fig.~\ref{fig:bupt_scenes}(a) shows the satellite view of the BUPT campus and the receiver regions used in this study. Fig.~\ref{fig:bupt_scenes}(b) provides a co-located local comparison, with the high-fidelity geometry shown on the left and the low-fidelity geometry shown on the right. Figs.~\ref{fig:bupt_scenes}(c) and (d) show the complete high-fidelity and low-fidelity scenes, respectively. The same coordinate axes, Tx location, and receiver-region annotations are used throughout the figure.

	The two profiles also differ substantially in geometric complexity. After the polygonal surfaces are triangulated by the RT scene loader, the low-fidelity scene contains 2,459 vertices and 1,313 triangular facets, whereas the high-fidelity scene contains 456,755 vertices and 193,421 triangular facets. The high-fidelity model therefore contains approximately 147 times as many triangular facets as the low-fidelity model. In addition, the surface materials in the high-fidelity scenario include seven types, such as concrete and metal, whereas all surfaces in the low-fidelity scenario are assigned concrete, with dielectric constants and other parameters derived from Recommendation ITU-R P.2040-4~\cite{ITU_P2040_4}. These differences establish the geometric and material-representation gap that the proposed network is designed to calibrate.
	
	\begin{table}[!t]
		\caption{Receiver Regions Used in the Simulation Study}
		\label{tab:receiver_regions}
		\centering
		\footnotesize
		\begin{tabular}{lccc}
			\toprule
			Region & $x$ range (m) & $y$ range (m) & Samples \\
			\midrule
			Train 1 & $[-240,-40]$ & $[-100,0]$ & 20,000 \\
			Train 2 & $[30,200]$ & $[50,100]$ & 17,000 \\
			Train 3 & $[-240,-80]$ & $[160,190]$ & 4,800 \\			
			Test 1 & $[-150,-110]$ & $[60,85]$ & 1,000 \\
			Test 2 & $[-80,-50]$ & $[65,95]$ & 900 \\
			\bottomrule
		\end{tabular}
	\end{table}
		
	Table~\ref{tab:receiver_regions} summarizes the five receiver regions using descriptive labels rather than scene-file identifiers. Train 1, Train 2, Train 3, Test 1, and Test 2 contain 20,000, 17,000, 4,800, 1,000, and 900 paired receiver samples, respectively. All receivers are placed at the same height, and the Tx position is fixed for every region.

	\subsubsection{Ray-Tracing Configuration}
	
	The paired channel data are generated using the RT engine employed in the DeepRT framework~\cite{Li2025}, with a shooting-and-bouncing-ray (SBR) solver. The initial ray directions are sampled over the sphere using Fibonacci-sphere sampling, and a hybrid path-search procedure is used to identify valid propagation paths. Line-of-sight propagation, specular reflection, penetration, and diffraction are enabled, whereas diffuse scattering is disabled in the present configuration. The same radio parameters, ray-launch budget, and path-search limits are used for the two profiles. Therefore, the controlled difference between a paired low-/high-fidelity sample is restricted to the geometric and material representations of the environment.
	
	The common RT parameters are summarized in Table~\ref{tab:rt_parameters}. The carrier frequency is $6.775$~GHz and the transmit power is $-7$~dBm. The Tx is located at $(-23.5,\,131.0,\,20.5)$~m, while every Rx is placed at a height of $2.5$~m. Both terminals use omnidirectional antennas. For each link, 300,000 rays are launched using Fibonacci-sphere sampling. The path-search configuration permits a maximum of four reflections, one penetration, and one diffraction interaction. These limits are shared by the low- and high-fidelity profiles so that their outputs differ only because of the paired scene representations.
	
	The network is trained with AdamW, an initial learning rate of $3\times10^{-4}$, a mini-batch size of $B=256$, cosine learning-rate annealing, and gradient-norm clipping at 2.0. The hidden dimension is $H=128$, the ordinary dropout rate is 0.1, and the maximum number of epochs is 250. Early stopping based on the validation loss is used during training. The training objective in \eqref{eq:total_loss_ch3} uses $\lambda_{P}=1.5$, $\lambda_{\sigma}=1$, $\lambda_{\mu}=1$, and $\lambda_{\mathrm{PDP}}=0.3$. For object selection, the top $N_{\mathrm{obj}}=32$ objects are retained from each geometry profile. The complete model and all ablation variants use these same settings. High-fidelity geometric groups are subject to structured masking during training and are unavailable at inference.
	
	\begin{table}[!t]
		\caption{Common Ray-Tracing Parameters for the Paired Profiles}
		\label{tab:rt_parameters}
		\centering
		\footnotesize
		\renewcommand{\arraystretch}{1.08}
		\begin{tabular}{@{}p{0.43\columnwidth}p{0.49\columnwidth}@{}}
			\toprule
			\textbf{Parameter} & \textbf{Value} \\
			\midrule
			Carrier frequency & 6.775~GHz \\
			Transmit power & $-7$~dBm \\
			Tx position & $(-23.5,\,131.0,\,20.5)$~m \\
			Rx height & 2.5~m \\
			Tx and Rx antennas & Omnidirectional \\
			Number of launched rays & 300,000 \\
			Ray-launch method & Fibonacci-sphere sampling \\
			Path-search method & Hybrid path search \\
			LoS propagation & Enabled \\
			Maximum reflections & 4 \\
			Maximum penetrations & 1 \\
			Maximum diffraction interactions & 1 \\
			Diffuse scattering & Disabled \\
			\bottomrule
		\end{tabular}
	\end{table}

	\subsection{Results and Discussion}
	\label{subsec:results_discussion}
	
	\subsubsection{Evaluation Metrics}
	
	The high-fidelity RT result is used as the reference throughout the evaluation. For receiver sample $i$ and scalar channel statistic $z$, the absolute errors of the low-fidelity RT baseline and the proposed model are defined as
	\begin{equation}
		e_{i,z}^{\mathrm{L}}
		=\left|z_i^{\mathrm{L}}-z_i^{\mathrm{H}}\right|,
		\qquad
		e_{i,z}^{\mathrm{M}}
		=\left|\hat{z}_i^{\mathrm{H}}-z_i^{\mathrm{H}}\right|,
		\label{eq:scalar_absolute_errors}
	\end{equation}
	where $z_i^{\mathrm{L}}$, $z_i^{\mathrm{H}}$, and $\hat{z}_i^{\mathrm{H}}$ denote the low-fidelity result, high-fidelity reference, and calibrated model output, respectively. The mean absolute error (MAE) is the average absolute error over the evaluation region, and the empirical cumulative distribution function (ECDF) reports the fraction of receiver locations whose absolute error does not exceed $\epsilon$.
	
	The scalar quantities evaluated below are total received power, RMS delay spread, and mean excess delay. All three quantities are obtained from the scalar prediction head after the corresponding inverse transformations. The normalized power-weighted excess-delay PDP is evaluated separately from the probability vector produced by the distribution head. For PDP comparison, the normalized high-fidelity distribution computed from the RT MPCs is
	\begin{equation}
		p_{i,b}^{\mathrm{H}}
		=\frac{\sum_{k\in\mathcal{K}_{i,W}^{\mathrm{H}}} P_{i,k}^{\mathrm{H}}
			\mathbb{I}\!\left(\tau_{i,k}^{\mathrm{ex},\mathrm{H}}\in\mathcal{B}_b\right)}
		{\sum_{k\in\mathcal{K}_{i,W}^{\mathrm{H}}} P_{i,k}^{\mathrm{H}}},
		\qquad
		\sum_{b=1}^{N_{\tau}}p_{i,b}^{\mathrm{H}}=1,
		\label{eq:eval_normalized_pdp}
	\end{equation}
	where $\mathbb{I}$ is the indicator function; PDP comparisons require positive in-window power for both RT profiles. The corresponding low-fidelity and model-reconstructed distributions are denoted by $\mathbf{p}_i^{\mathrm{L}}$ and $\widehat{\mathbf{p}}_i^{\mathrm{H}}$, respectively. For two normalized PDPs $\mathbf{p}$ and $\mathbf{q}$, their $L_1$ distance and Jensen--Shannon (JS) divergence are
	\begin{align}
		D_{1}(\mathbf{p},\mathbf{q})
		&=\sum_{b=1}^{N_{\tau}}|p_b-q_b|,\\
		D_{\mathrm{JS}}(\mathbf{p},\mathbf{q})
		&=\frac{1}{2}\sum_{b=1}^{N_{\tau}}p_b
		\ln\!\frac{p_b}{m_b}
		+\frac{1}{2}\sum_{b=1}^{N_{\tau}}q_b
		\ln\!\frac{q_b}{m_b},
		\label{eq:pdp_l1_js}\\
		m_b&=\frac{p_b+q_b}{2}.
	\end{align}
	Natural logarithms are used, with zero-probability terms interpreted by continuity as zero. Lower values of both measures indicate closer agreement, with $0\leq D_1\leq2$ and $0\leq D_{\mathrm{JS}}\leq\ln2$.

	\begin{figure}[!t]
		\centering
		\includegraphics[width=\columnwidth]{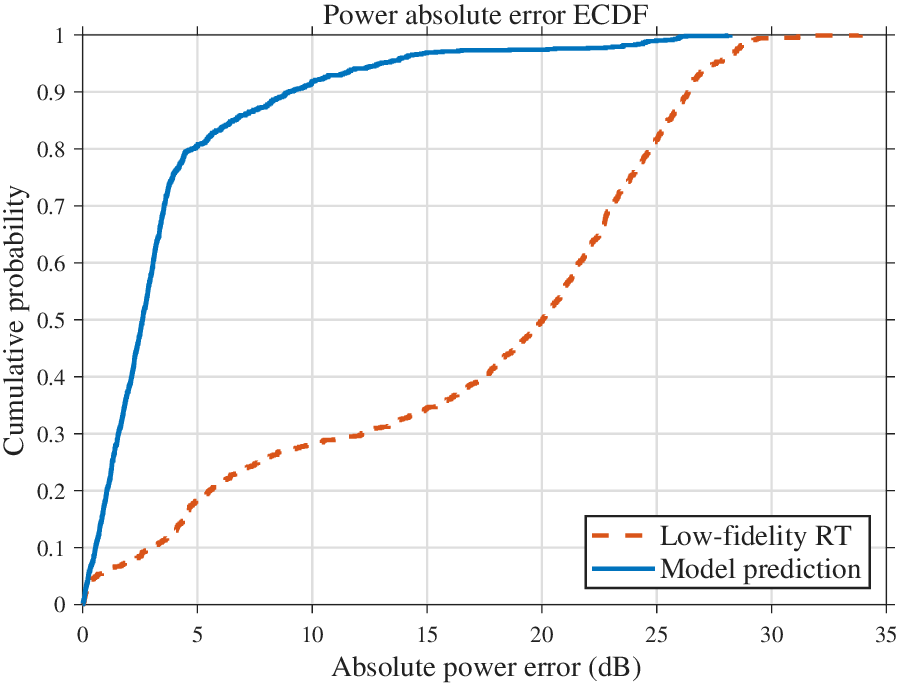}
		\caption{Empirical cumulative distribution functions (ECDFs) of the absolute total-received-power error in Test 1, evaluated with respect to the high-fidelity RT reference. The proposed model reduces the mean absolute error (MAE) from $16.80$~dB for low-fidelity RT to $3.88$~dB.}
		\label{fig:test1_power_ecdf}
	\end{figure}
	
	\begin{figure}[!t]
		\centering
		\includegraphics[width=\columnwidth]{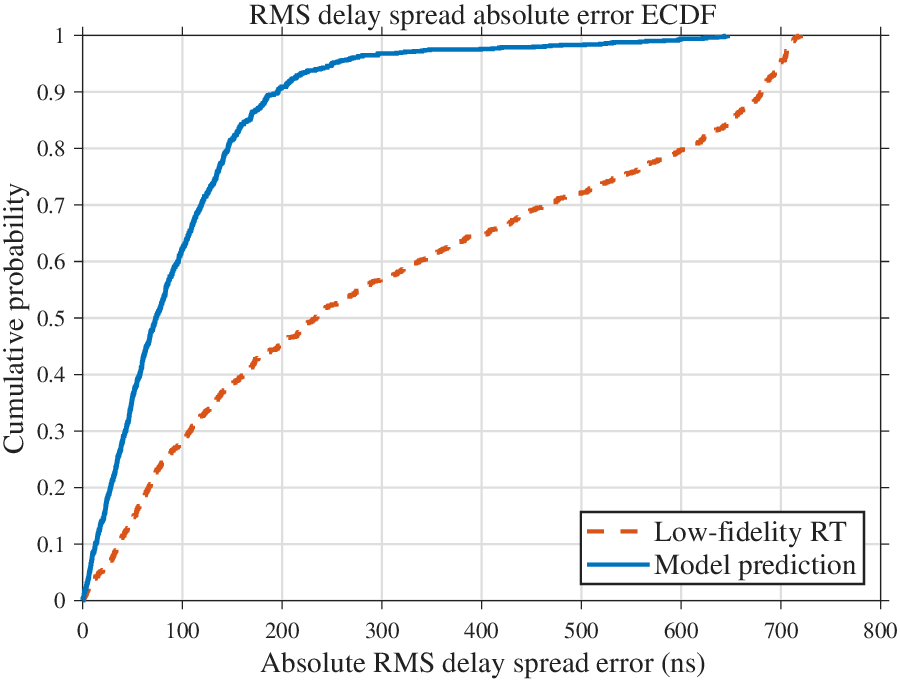}
		\caption{ECDFs of the absolute RMS delay-spread error in Test 1 relative to the high-fidelity RT reference. The MAE is reduced from $304.89$~ns for low-fidelity RT to $98.19$~ns after calibration.}
		\label{fig:test1_rms_ecdf}
	\end{figure}
	
	\begin{figure}[!t]
		\centering
		\includegraphics[width=\columnwidth]{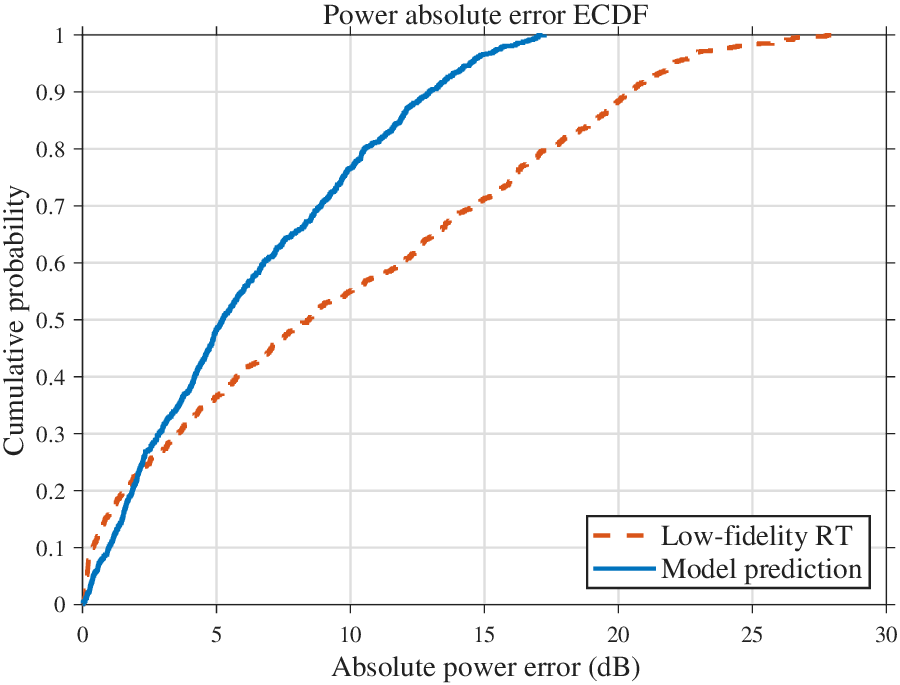}
		\caption{ECDFs of the absolute total-received-power error in the LoS-dominated Test 2 region. Relative to high-fidelity RT, the proposed model achieves an MAE of $6.26$~dB, compared with $9.64$~dB for low-fidelity RT.}
		\label{fig:test2_power_ecdf}
	\end{figure}
	
	\begin{figure}[!t]
		\centering
		\includegraphics[width=\columnwidth]{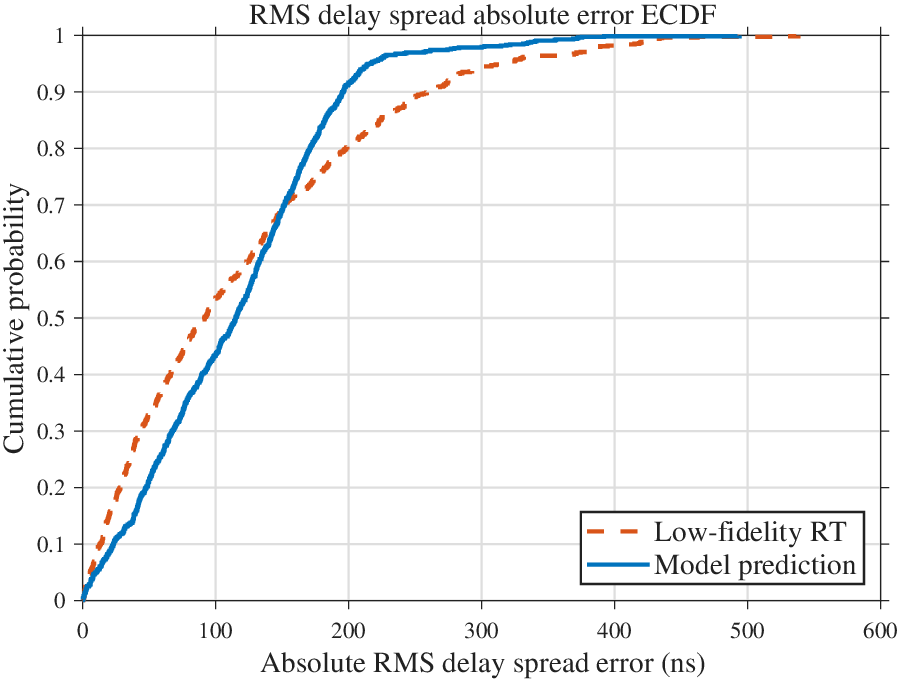}
		\caption{ECDFs of the absolute RMS delay-spread error in Test 2. The model and low-fidelity RT yield similar error distributions, with MAEs of $117.37$~ns and $114.36$~ns, respectively; the proposed calibration is therefore slightly worse for this metric.}
		\label{fig:test2_rms_ecdf}
	\end{figure}
	
\begin{figure*}[!t]
		\centering
		\newcommand{\spatialrow}[6]{%
			\noindent\begin{minipage}[c]{0.04\textwidth}\centering\rotatebox{90}{\footnotesize #1}\end{minipage}\hfill%
			\begin{minipage}[c]{0.18\textwidth}\centering\includegraphics[width=\linewidth]{#2}\end{minipage}\hfill%
			\begin{minipage}[c]{0.18\textwidth}\centering\includegraphics[width=\linewidth]{#3}\end{minipage}\hfill%
			\begin{minipage}[c]{0.18\textwidth}\centering\includegraphics[width=\linewidth]{#4}\end{minipage}\hfill%
			\begin{minipage}[c]{0.18\textwidth}\centering\includegraphics[width=\linewidth]{#5}\end{minipage}\hfill%
			\begin{minipage}[c]{0.18\textwidth}\centering\includegraphics[width=\linewidth]{#6}\end{minipage}\par\vspace{2pt}}
		\spatialrow{Power}{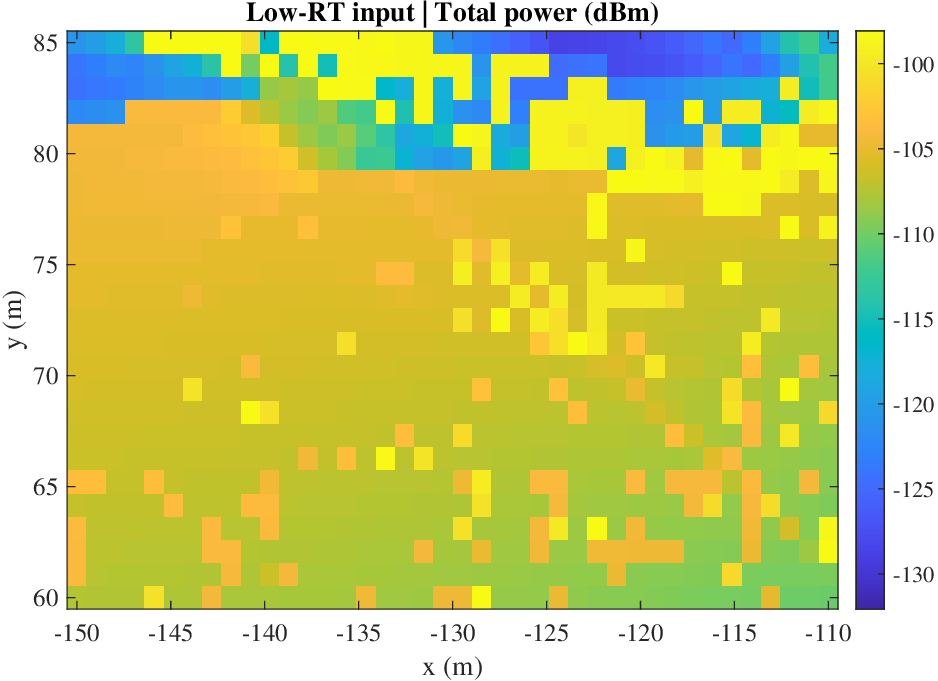}{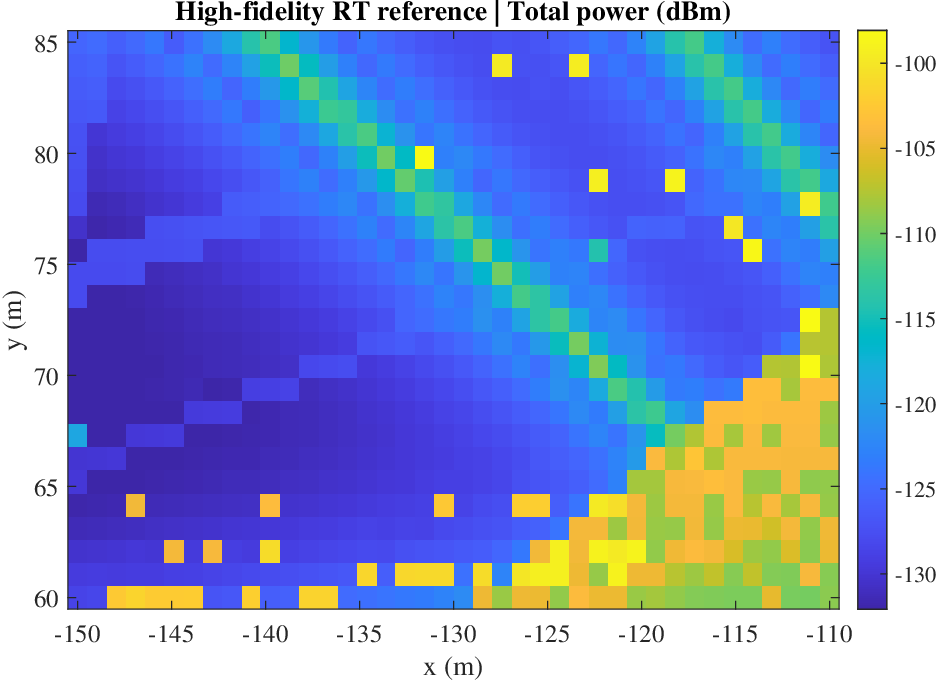}{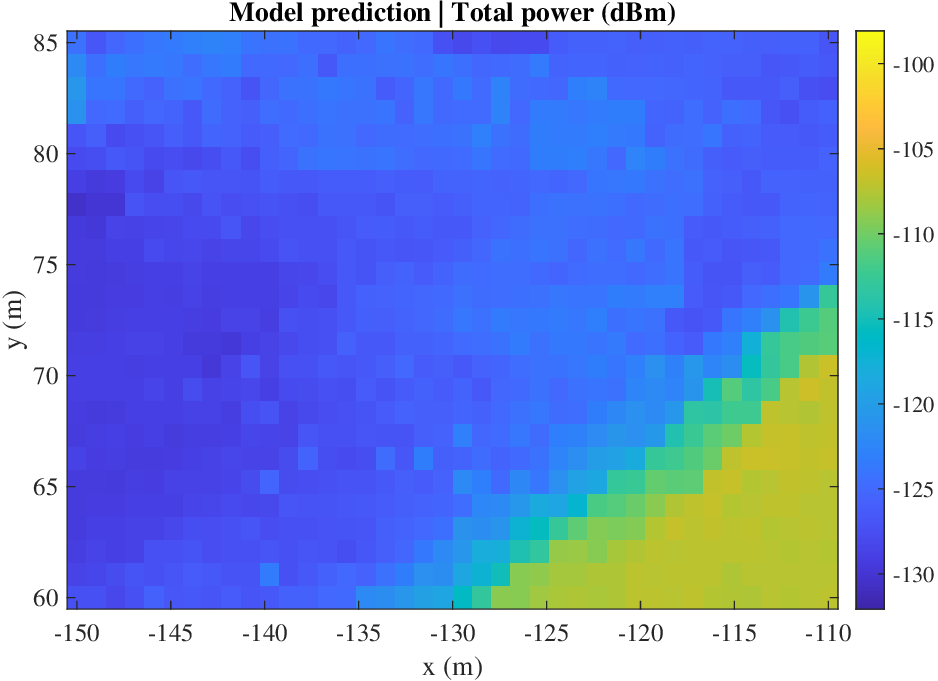}{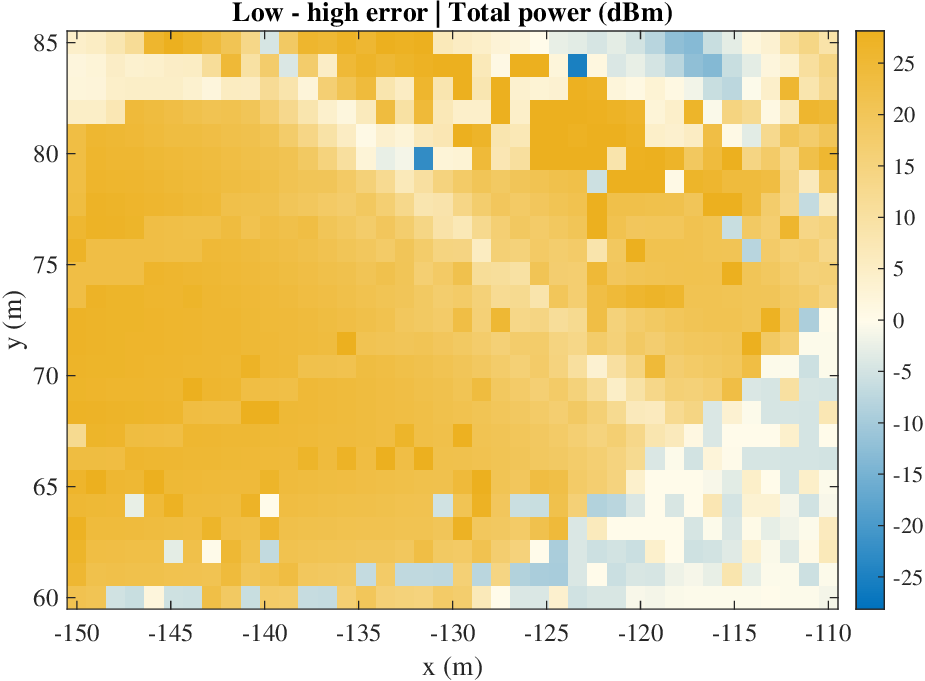}{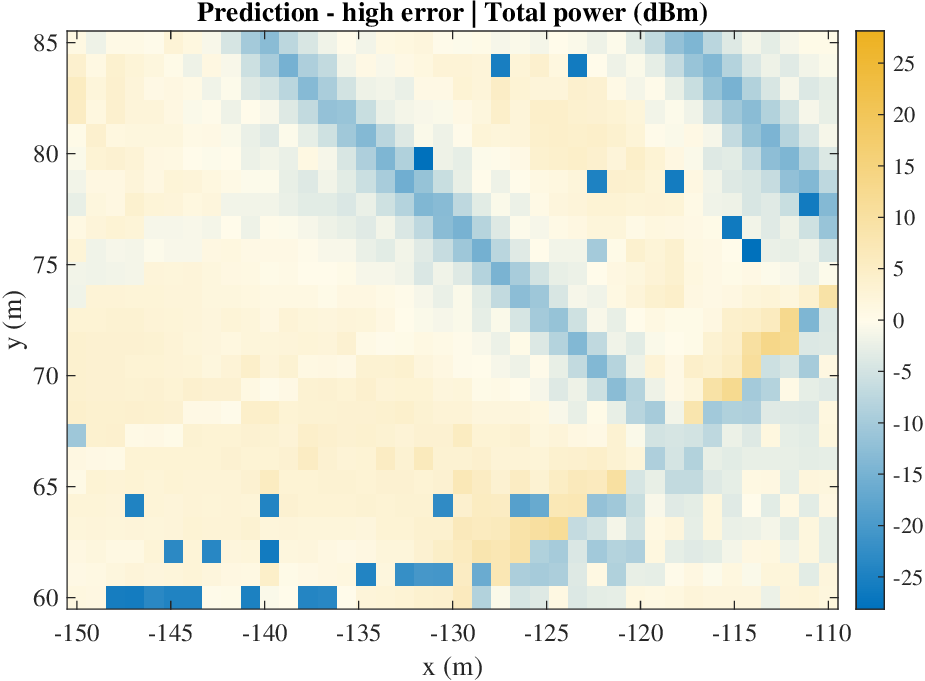}
		\spatialrow{RMS delay spread}{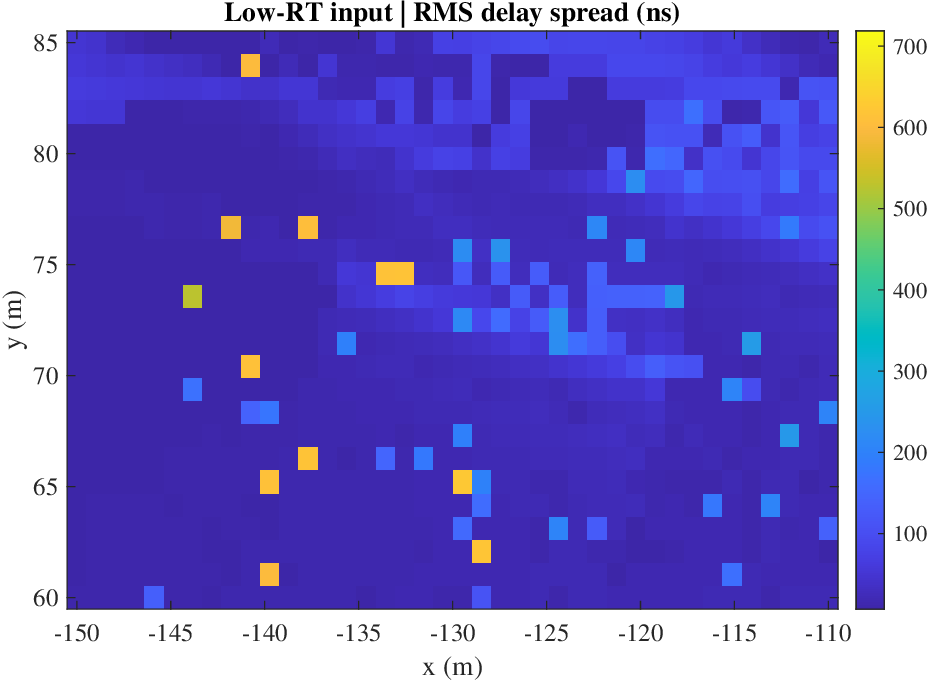}{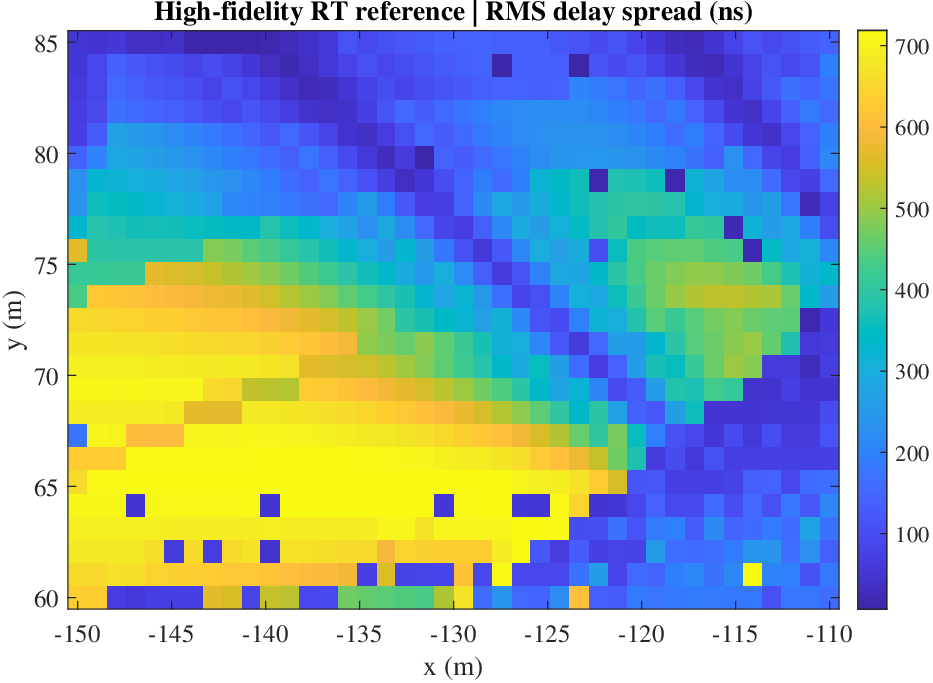}{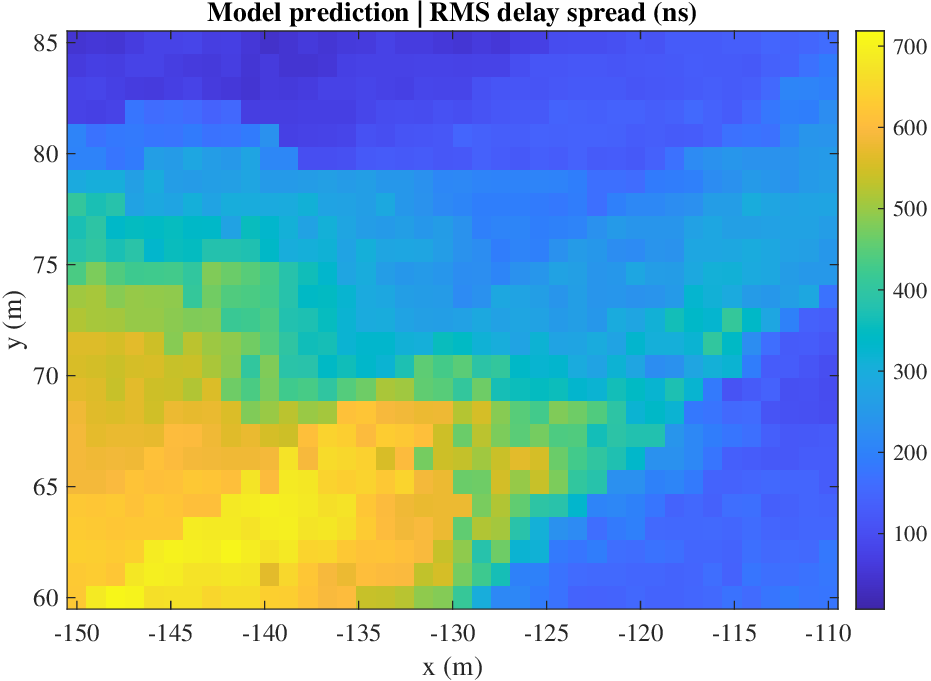}{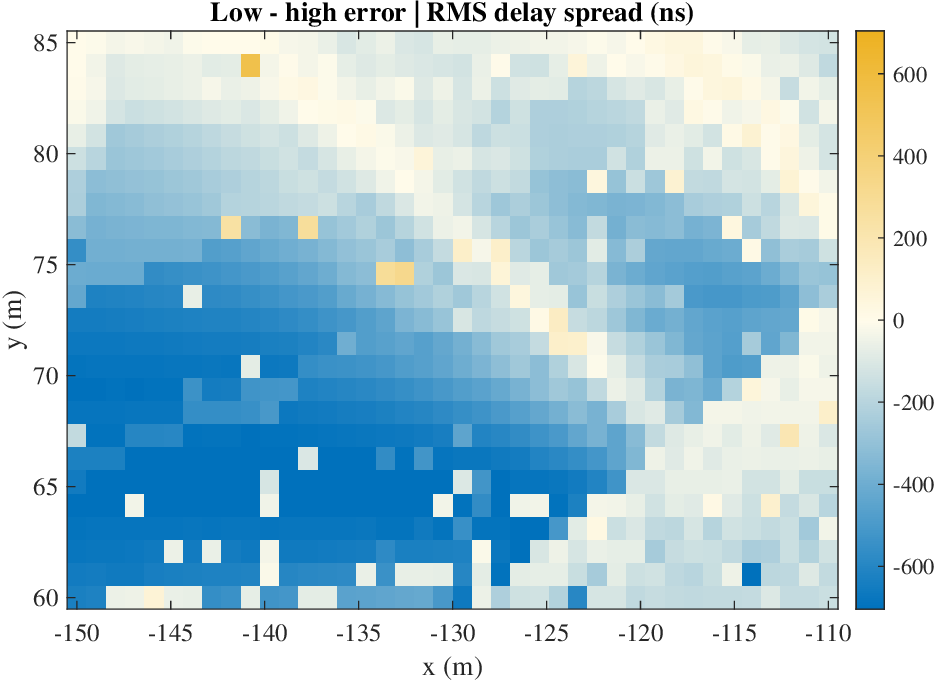}{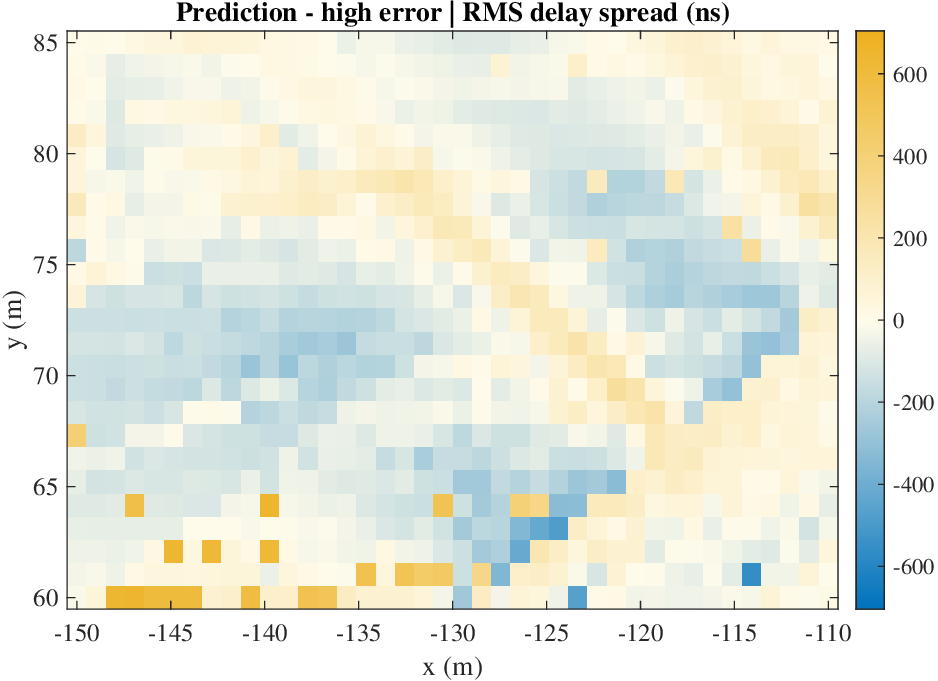}
		\spatialrow{Mean excess delay}{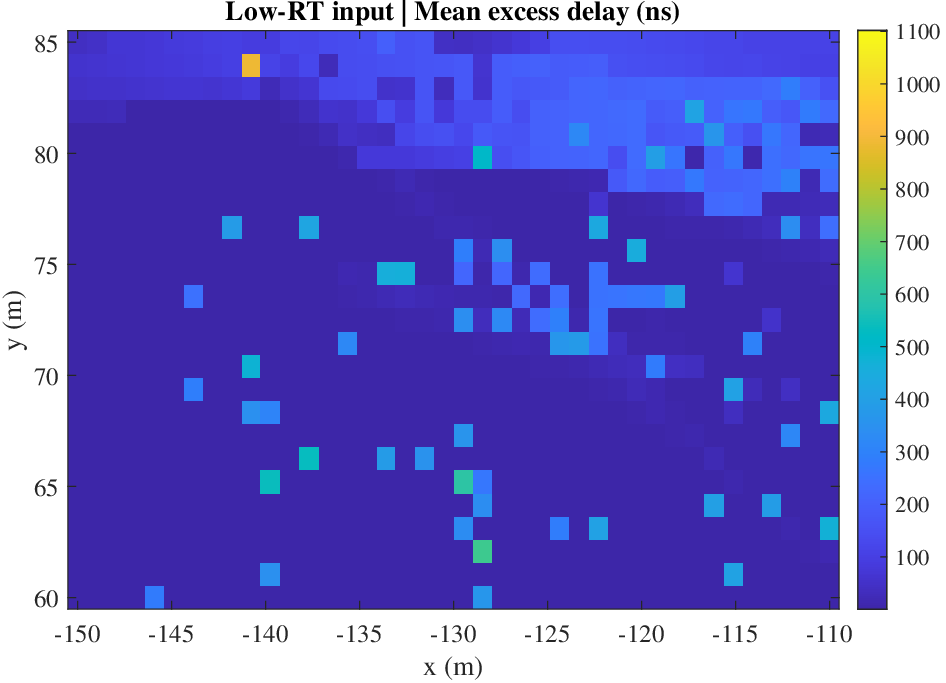}{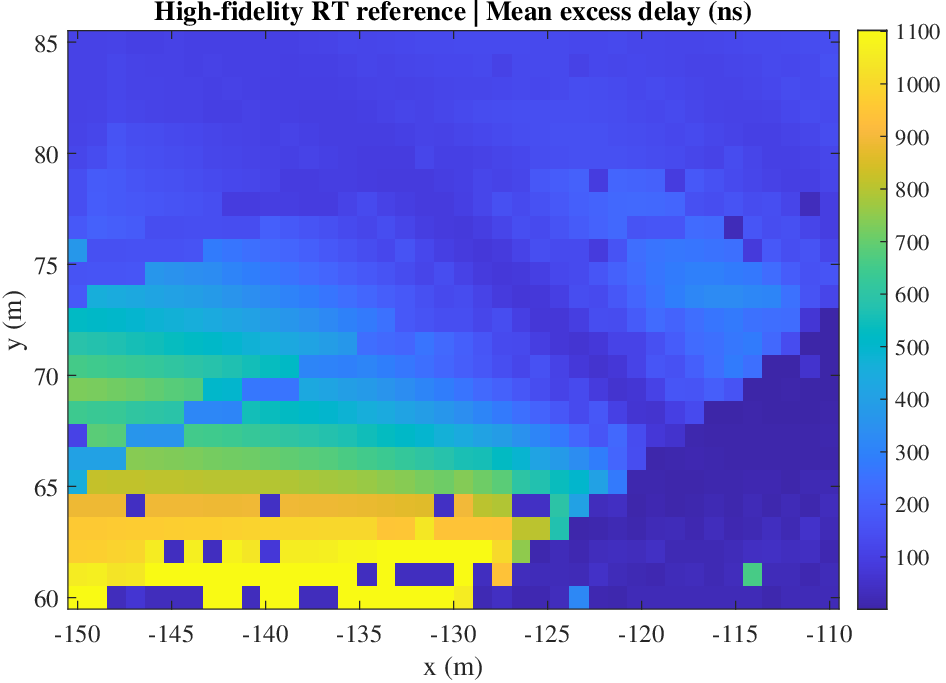}{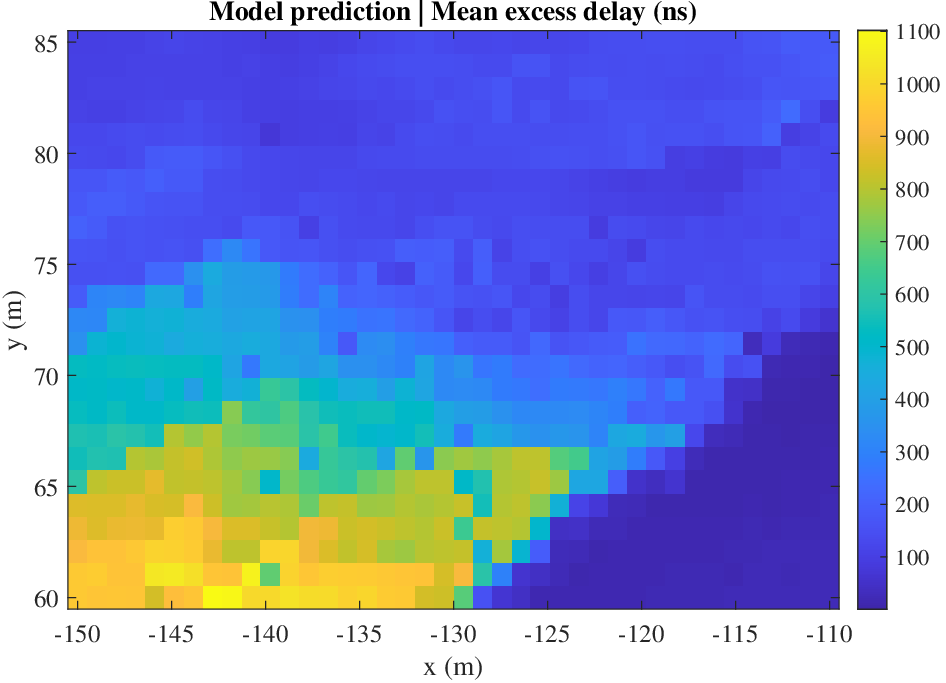}{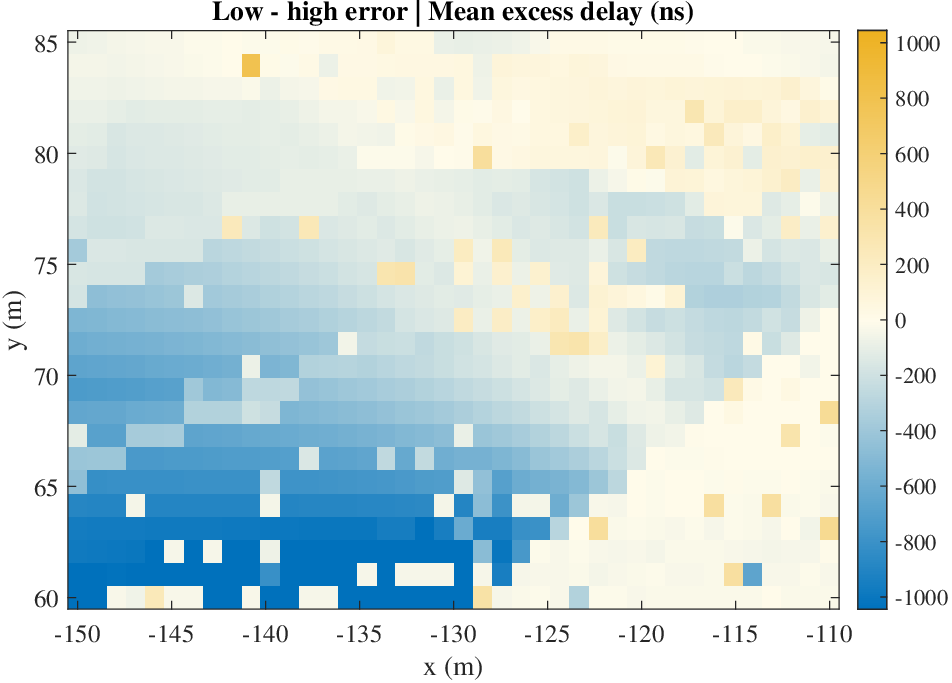}{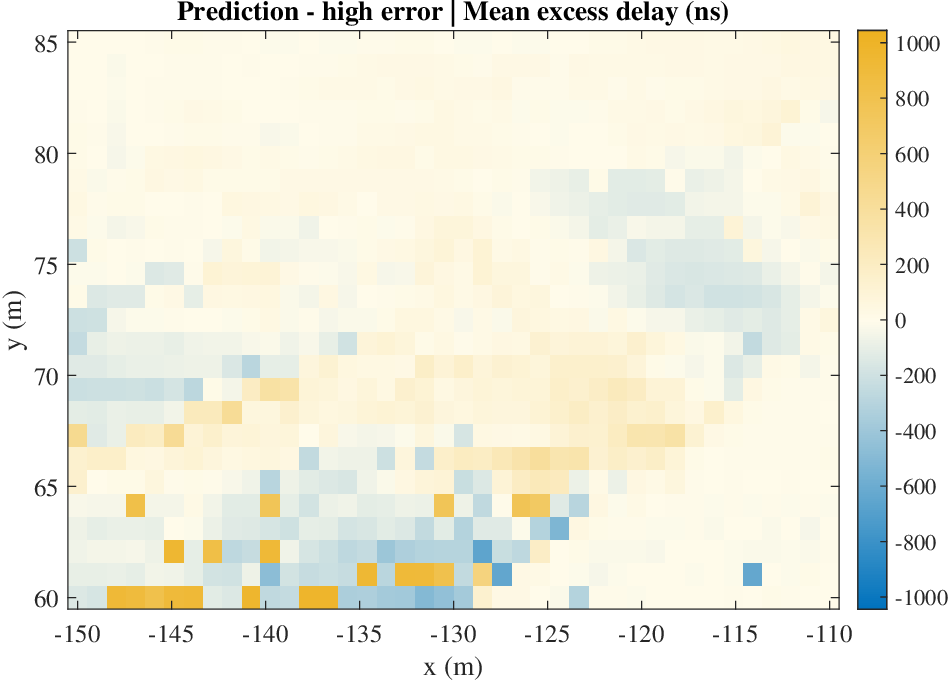}
		\vspace{-1pt}
		\noindent\begin{minipage}{0.04\textwidth}\end{minipage}\hfill%
		\begin{minipage}{0.24\textwidth}\centering\footnotesize (a) Low-RT\end{minipage}\hfill%
		\begin{minipage}{0.12\textwidth}\centering\footnotesize (b) High-RT\end{minipage}\hfill%
		\begin{minipage}{0.20\textwidth}\centering\footnotesize (c) Prediction\end{minipage}\hfill%
		\begin{minipage}{0.16\textwidth}\centering\footnotesize (d) Low-RT error\end{minipage}\hfill%
		\begin{minipage}{0.18\textwidth}\centering\footnotesize (e) Prediction error\end{minipage}
		\caption{Spatial comparison over Test 1. Rows are labeled by channel statistic; columns (a)--(e) show the low-fidelity RT input, high-fidelity RT reference, model prediction, signed low-fidelity error $z^{\mathrm{L}}-z^{\mathrm{H}}$, and signed prediction error $\hat{z}^{\mathrm{H}}-z^{\mathrm{H}}$, respectively. Mean excess delay is referenced to the geometric Tx--Rx propagation delay. Within each row, the first three panels use a common value scale and the last two use a common signed-error scale.}
		\label{fig:test1_spatial_maps}
	\end{figure*}

	\subsubsection{Error Distributions in Test 1 and Test 2}
	
	Fig.~\ref{fig:test1_power_ecdf} shows that the proposed calibration reduces the Test 1 power MAE from $16.80$~dB to $3.88$~dB. The ECDF separation indicates that the improvement is spatially widespread rather than being produced by a small number of favorable receiver locations.

	Fig.~\ref{fig:test1_rms_ecdf} shows that the RMS delay-spread MAE decreases from $304.89$~ns to $98.19$~ns. The learned correction therefore recovers a substantial portion of the delay-dispersion discrepancy caused by the simplified scene representation, rather than improving received power alone.

	Test 2 is a more difficult, LoS-dominated spatial-generalization case in which the two fidelity profiles already have a smaller discrepancy. Nevertheless, Fig.~\ref{fig:test2_power_ecdf} shows that calibration reduces the power MAE from $9.64$~dB to $6.26$~dB.

	In the LoS-dominated Test 2 region, the calibrated RMS delay-spread error in Fig.~\ref{fig:test2_rms_ecdf} is comparable to the baseline ($117.37$~ns versus $114.36$~ns), indicating that the calibration gain is propagation-regime dependent. After the dominant direct component is accounted for, comparatively weak non-LoS components control the second-order delay moment, so small reconstruction differences can change RMS delay spread without materially altering the dominant received-power behavior.

	\subsubsection{Spatial Reconstruction in Test 1}
	
	The aggregate errors do not reveal whether the correction preserves the spatial organization of the channel. Fig.~\ref{fig:test1_spatial_maps} therefore compares the low-fidelity input, high-fidelity reference, model output, and two signed error fields over Test 1; the last two columns show $z_i^{\mathrm{L}}-z_i^{\mathrm{H}}$ and $\hat{z}_i^{\mathrm{H}}-z_i^{\mathrm{H}}$, respectively. The top row shows that the low-fidelity total-power field contains broad systematic offsets and misses several localized variations in the high-fidelity reference. The calibrated map follows the reference more closely, and its residual field has a substantially smaller magnitude, consistent with the reduction in power MAE from $16.80$~dB to $3.88$~dB.
	
	The middle row demonstrates a corresponding recovery of the large-scale RMS delay-spread structure. Although localized residuals remain near abrupt spatial transitions, the model reconstructs the principal high-delay regions and reduces the MAE from $304.89$~ns to $98.19$~ns. The bottom row evaluates the mean excess delay, which suppresses the common geometric-delay offset and emphasizes the relative organization of delayed multipath energy. For this statistic, the MAE decreases from $268.70$~ns for low-fidelity RT to $83.44$~ns for the model. Taken together, the three rows show that the network learns a spatially structured calibration rather than a region-wide scalar bias correction.

	\subsubsection{Representative Excess-Delay PDPs}
	
\begin{figure}[!t]
		\centering
		\includegraphics[width=\columnwidth]{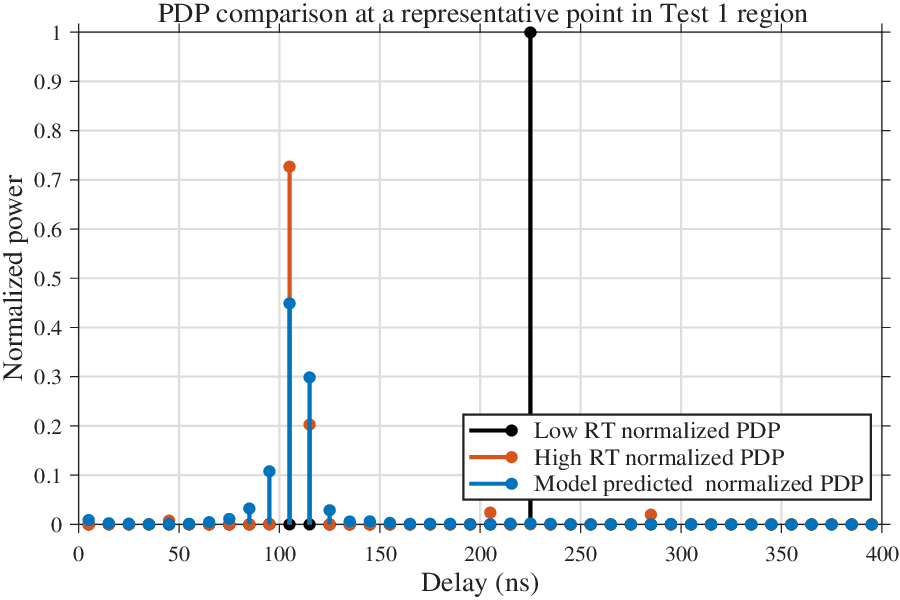}
		\caption{Normalized excess-delay PDPs at a representative receiver in Test 1. Delays are referenced to the geometric Tx--Rx propagation delay and accumulated in 10-ns bins. The model-reconstructed PDP recovers the dominant high-fidelity cluster near $100$~ns to $120$~ns, whereas the low-fidelity RT energy is concentrated near $220$~ns to $230$~ns.}
		\label{fig:test1_pdp}
	\end{figure}

\begin{figure}[!t]
		\centering
		\includegraphics[width=\columnwidth]{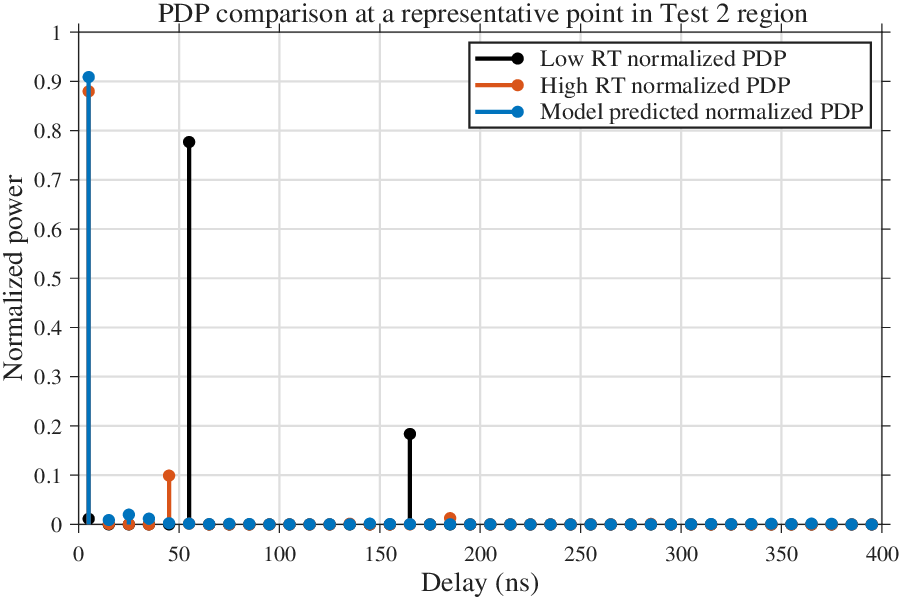}
		\caption{Normalized excess-delay PDPs at a representative receiver in the LoS-dominated Test 2 region, using the geometric Tx--Rx propagation delay as the reference and 10-ns bins. The model reconstructs the dominant high-fidelity component in the $0$~ns to $10$~ns bin, while the low-fidelity RT profile places most of its energy at larger excess delays.}
		\label{fig:test2_pdp}
	\end{figure}

	The scalar statistics above compress the multipath-delay structure into a few moments. To examine whether the calibrated output also reconstructs the distribution of delayed energy, Figs.~\ref{fig:test1_pdp} and \ref{fig:test2_pdp} present normalized excess-delay PDPs at representative receiver locations. All three distributions are evaluated using 10-ns bins after subtracting the geometric Tx--Rx propagation delay.
	
	At the representative Test 1 point in Fig.~\ref{fig:test1_pdp}, the low-fidelity RT result concentrates $99.92\%$ of its normalized power in the $220$~ns to $230$~ns bin. In contrast, the high-fidelity reference places $72.67\%$ and $20.28\%$ in the $100$~ns to $110$~ns and $110$~ns to $120$~ns bins, respectively. The reconstructed PDP relocates most of the energy to this earlier cluster, assigning $44.90\%$ and $29.87\%$ to the same two bins, with a further $10.80\%$ in the adjacent $90$~ns to $100$~ns bin. Over the complete delay range contained in the data, the model reduces the $L_1$ distance from $2.00$ to $0.67$ and the JS divergence from $0.69$ to $0.12$. The remaining broadening around the dominant cluster explains why the reconstructed PDP is substantially improved but not identical to the high-fidelity reference.
	
	Fig.~\ref{fig:test2_pdp} illustrates a different propagation regime. At this LoS-dominated Test 2 point, the high-fidelity reference assigns $87.96\%$ of the normalized power to the $0$~ns to $10$~ns excess-delay bin, and the reconstructed PDP assigns $90.87\%$ to the same bin. The low-fidelity RT result instead places $77.67\%$ and $18.38\%$ in the $50$~ns to $60$~ns and $160$~ns to $170$~ns bins. Consequently, reconstruction reduces the $L_1$ distance from $1.97$ to $0.23$ and the JS divergence from $0.66$ to $0.06$. The near-zero excess delay indicates that the path length is close to the geometric Tx--Rx distance after removal of the common propagation delay.
	
	This local distributional improvement coexists with the comparable region-wide RMS delay-spread errors in Fig.~\ref{fig:test2_rms_ecdf}. A representative PDP and an aggregate second-order statistic characterize different aspects of performance: the profile can be locally accurate while small residual errors across many LoS links still produce a slightly higher regional RMS delay-spread MAE.
	
\subsection{Ablation Study}
\label{subsec:ablation_study}

\subsubsection{Ablation Protocol and Metrics}

To isolate the contributions of the principal feature branches and the link-query cross-attention module, four reduced variants are compared with the complete model on Test 1. In the \emph{without path features} variant, the Path Encoder is removed and the path representation is excluded from both the link query and the final fusion. In the \emph{without object attention} variant, the Object Projection and Link-Query Cross-Attention modules are removed completely; only the global, path, and scene representations participate in fusion. In the \emph{without global features} variant, the Global Encoder output is removed from both query construction and final fusion, so the link query is formed from the path representation alone. Finally, the \emph{without path features and object attention} variant retains only the global and scene branches.

All five configurations use the training protocol specified in Section~\ref{subsec:simulation_setup}, including identical training samples, Test 1 receivers, model dimensions, optimization settings, loss weights, maximum epoch, and early-stopping rule. The comparison changes only the specified architectural components.

The ablation metrics are MAE, median absolute error (the 50th percentile), and the standard Pearson correlation between the prediction and high-fidelity reference over all Test 1 receivers. MAE measures average error magnitude, the median represents the typical receiver, and Pearson correlation measures whether the predicted spatial ordering follows the reference.

\subsubsection{Quantitative Ablation Results}

\begin{table*}[!t]
	\caption{Quantitative Ablation Results on Test 1}
	\label{tab:test1_ablation}
	\centering
	\footnotesize
	\setlength{\tabcolsep}{4pt}
	\begin{tabular}{@{}lcccccc@{}}
		\toprule
		Variant & Power MAE (dB) & RMS MAE (ns) & Power corr. & RMS corr.
		& Power $|e|$ p50 (dB) & RMS $|e|$ p50 (ns) \\
		\midrule
		Without path features
		& 4.16 & 104.06 & 0.75 & 0.81 & 3.21 & 82.06 \\
		Without object attention
		& 4.28 & 103.47 & 0.72 & 0.81 & 3.07 & 77.66 \\
		Without global features
		& 3.95 & \textbf{96.14} & \textbf{0.76} & \textbf{0.82} & 2.75 & 76.51 \\
		Without path features and object attention
		& 4.55 & 105.17 & 0.72 & 0.78 & 3.74 & 75.36 \\
		\midrule
		Complete model
		& \textbf{3.88} & 98.19 & 0.75 & 0.82 & \textbf{2.61} & \textbf{73.55} \\
		\bottomrule
	\end{tabular}
	\vspace{2pt}
	
	\begin{minipage}{0.98\textwidth}
		\footnotesize
		\textit{Notes:} All errors are computed with respect to the high-fidelity RT reference. ``corr.'' denotes the Pearson correlation between the prediction and reference over all Test 1 receivers, and p50 denotes the median absolute error. Boldface identifies the best value using the unrounded results. All variants use an identical training and evaluation protocol.
	\end{minipage}
\end{table*}

Table~\ref{tab:test1_ablation} summarizes the controlled comparison. For total received power, the complete model achieves the lowest MAE of $3.88$~dB and the lowest median error of $2.61$~dB. Removing path features increases these errors to $4.16$~dB and $3.21$~dB, respectively. Directly removing object attention increases them to $4.28$~dB and $3.07$~dB. When both path features and object attention are removed, the power MAE reaches $4.55$~dB and the median error reaches $3.74$~dB, which are the largest values among the five variants. The larger degradation of the combined ablation suggests that explicit low-fidelity path evidence and link-conditioned object context provide complementary information for power calibration.

The RMS delay-spread results exhibit a less uniform ranking. The complete model gives an MAE of $98.19$~ns and the best median error of $73.55$~ns. Removing path features or object attention increases the MAE to $104.06$~ns and $103.47$~ns, respectively, confirming that both branches contribute to delay-dispersion calibration. The combined removal produces the largest RMS MAE of $105.17$~ns and the lowest RMS correlation of $0.78$. Its median error remains comparatively small at $75.36$~ns, showing that MAE and median error can rank the variants differently.

Removing global features yields the lowest RMS MAE of $96.14$~ns and the highest power and RMS correlations, $0.76$ and $0.82$, while its power MAE of $3.95$~dB remains close to that of the complete model. However, the complete model retains lower median errors for both targets. The global branch therefore does not improve every aggregate metric. Instead, the results indicate partial redundancy among global, path, and scene information, together with a tradeoff between mean and median absolute errors. Accordingly, the ablation evidence supports the clear contributions of the path and object-attention branches, while the incremental role of the global branch is more metric dependent.

\begin{figure}[!t]
	\centering
	\includegraphics[width=\columnwidth]{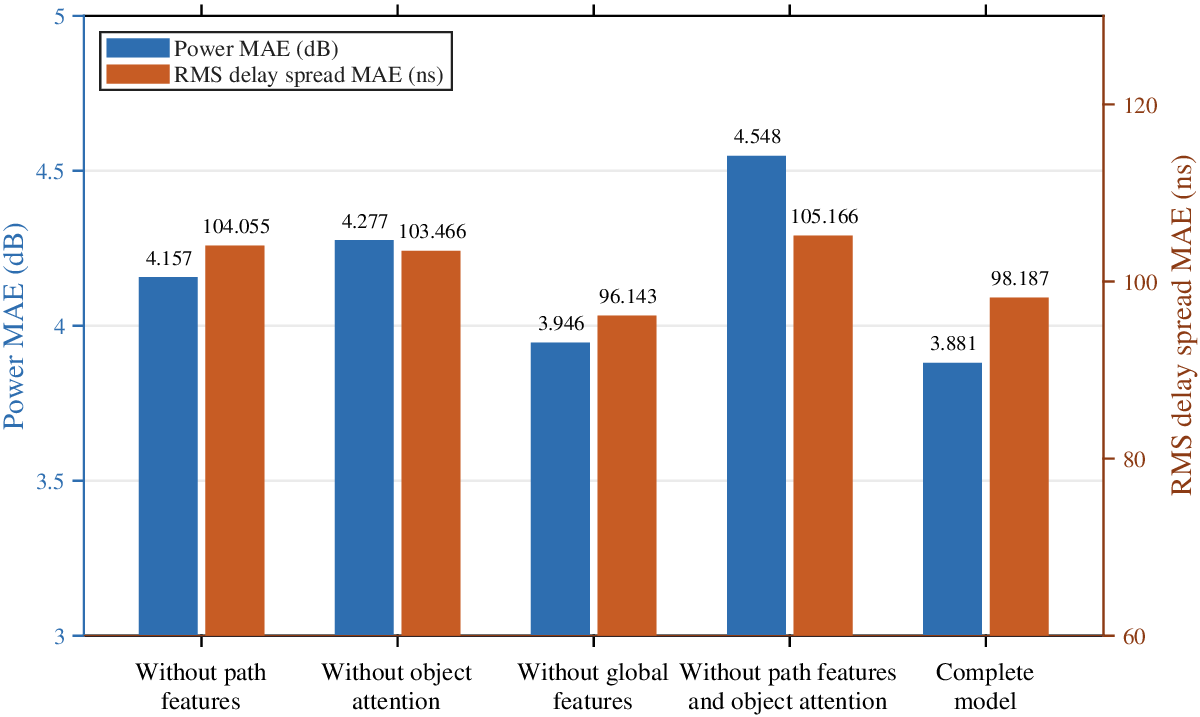}
	\caption{MAEs of total received power and RMS delay spread for the five ablation configurations on Test 1, evaluated against the high-fidelity RT reference. Blue bars use the left axis (dB), and orange bars use the right axis (ns). Lower values indicate smaller errors.}
	\label{fig:test1_ablation_mae}
\end{figure}

Fig.~\ref{fig:test1_ablation_mae} compares the power and RMS delay-spread MAEs of the five configurations. The complete model achieves the lowest power MAE, while the variant without global features gives the lowest RMS delay-spread MAE. Removing both path features and object attention produces the largest MAEs for both metrics, supporting their complementary roles in calibration.

\subsubsection{Spatial Effects of Path and Object-Attention Ablation}

\begin{figure}[!t]
	\centering
	\subfloat[Without object attention: power]{
		\includegraphics[width=0.45\columnwidth]{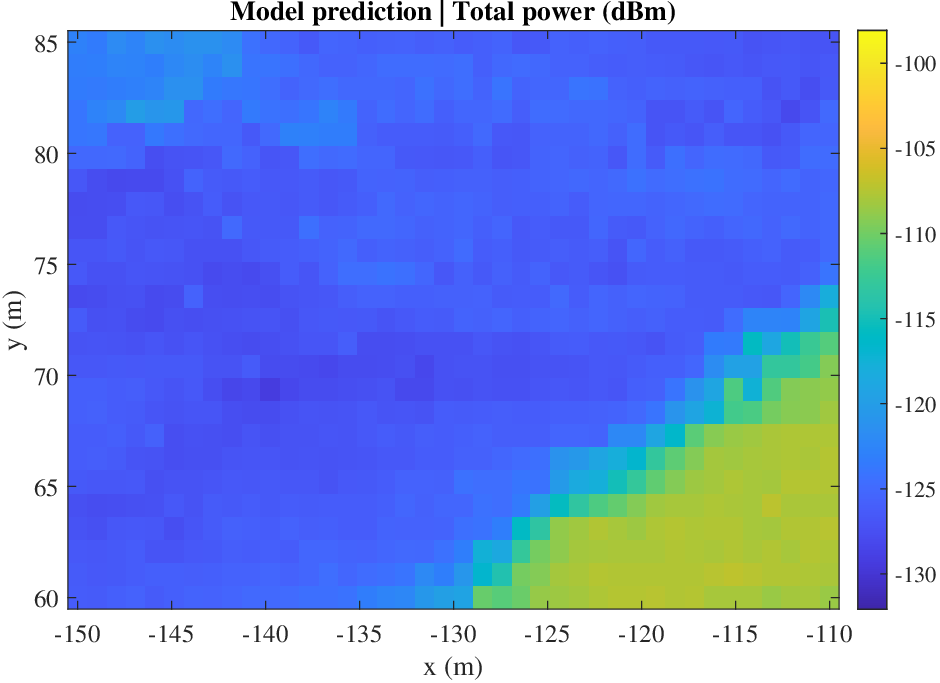}}
	\hfill
	\subfloat[Without path features: power]{
		\includegraphics[width=0.45\columnwidth]{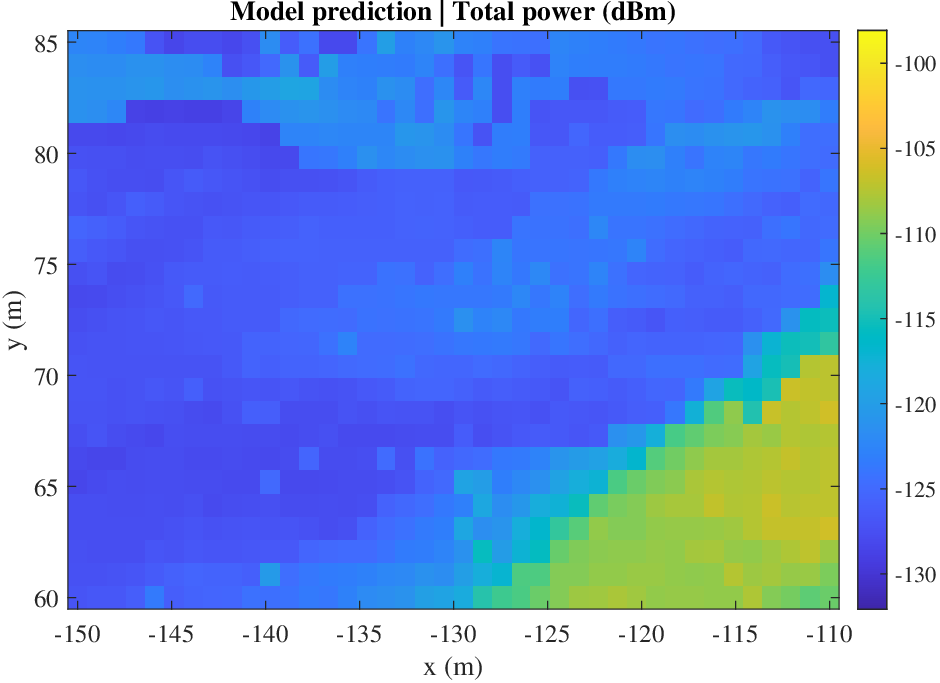}}\\
	\subfloat[Without object attention: RMS delay spread]{
		\includegraphics[width=0.45\columnwidth]{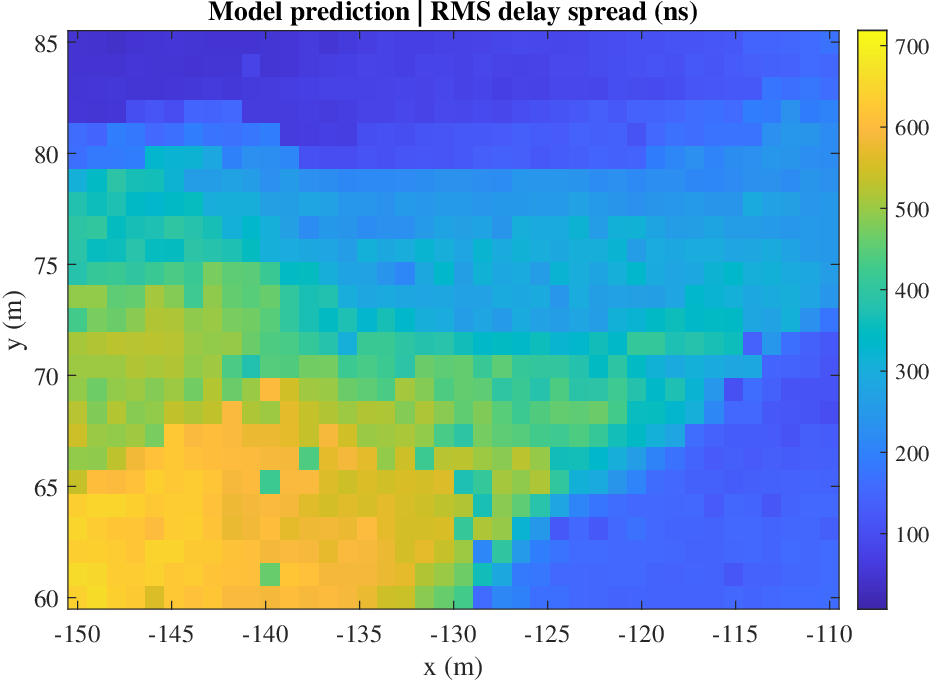}}
	\hfill
	\subfloat[Without path features: RMS delay spread]{
		\includegraphics[width=0.45\columnwidth]{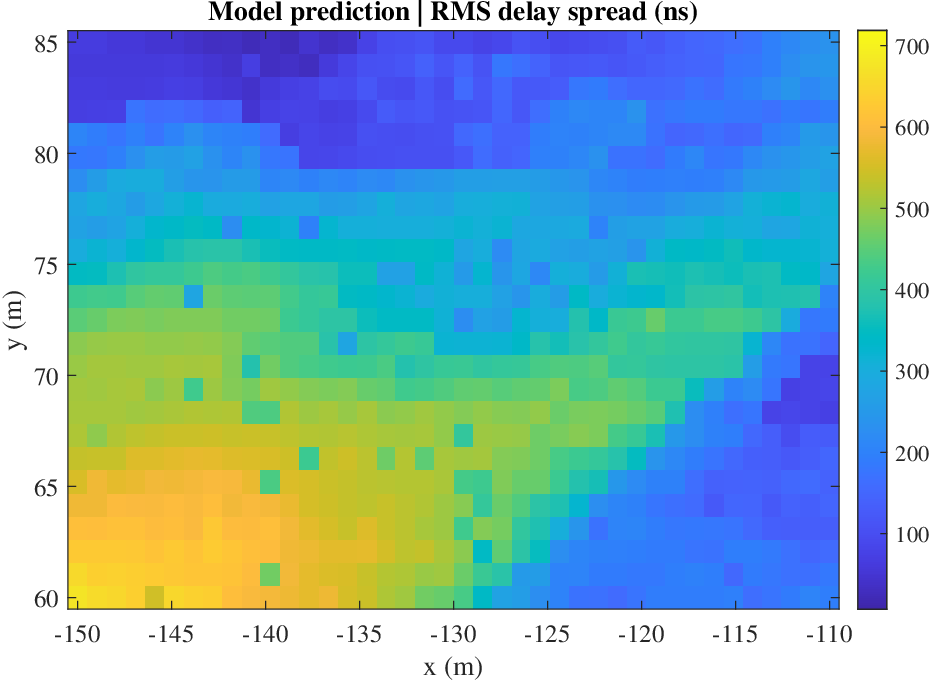}}
	\caption{Spatial predictions of two ablated models over Test 1. The top and bottom rows show total received power and RMS delay spread, respectively. The left column removes the Object Projection and Link-Query Cross-Attention branch, while the right column removes the Path Encoder and excludes the path representation from query construction and fusion. The coordinate ranges and metric-specific color scales are shared within each row.}
	\label{fig:test1_ablation_spatial}
\end{figure}

Fig.~\ref{fig:test1_ablation_spatial} provides a spatial view of the two ablations that most directly affect link-specific propagation evidence. The top and bottom rows show total received power and RMS delay spread, respectively; the left column removes object attention, and the right column removes path features. Both reduced models preserve the broad spatial trend because the global and scene branches remain active. Nevertheless, their local transitions and regional amplitudes differ from the high-fidelity reference shown in Fig.~\ref{fig:test1_spatial_maps}. Removing object attention eliminates the mechanism that selects environmental-object evidence conditioned on the current link, while removing path features discards the explicit low-fidelity MPC-set representation from both query construction and fusion. The resulting spatial distortions are consistent with the larger power and RMS errors reported in Table~\ref{tab:test1_ablation}.

Broad visual agreement can coexist with localized prediction errors. Together, the spatial maps and quantitative metrics show that path tokens and link-query object attention provide complementary information for calibration.

\section{Conclusion}
\label{sec:conclusion}

This paper presented DeepRT-Cal, a geometry-aware framework for calibrating low-fidelity RT results toward twin channel statistics obtained from detailed-geometry RT. By formulating cross-fidelity calibration as a set-to-statistics learning problem, the proposed method avoids imposing an unreliable one-to-one correspondence between low- and high-fidelity MPCs. DeepRT-Cal jointly encodes global link descriptors, the variable-cardinality low-fidelity MPC set, propagation-corridor scene features, and link-centered environmental objects. Its link-query cross-attention mechanism retrieves object context conditioned on the current link, and the fused representation reconstructs high-fidelity scalar channel statistics and normalized excess-delay distributions.

Outdoor experiments showed that the gains over direct low-fidelity RT depend on the propagation regime. In Test~1, DeepRT-Cal reduced the MAEs of total received power, RMS delay spread, and mean excess delay from $16.80$~dB, $304.89$~ns, and $268.70$~ns to $3.88$~dB, $98.19$~ns, and $83.44$~ns, respectively. At the representative Test~1 receiver, the JS divergence of the normalized excess-delay PDP decreased from $0.69$ to $0.12$. In the LoS-dominated Test~2 region, the power MAE decreased from $9.64$~dB to $6.26$~dB, and the representative PDP JS divergence decreased from $0.66$ to $0.06$. The RMS delay-spread MAE changed from $114.36$~ns to $117.37$~ns, confirming that the calibration gain is propagation-regime dependent. The ablation results further showed that low-fidelity path features and link-conditioned object attention provide complementary calibration evidence, whereas the contribution of the global branch is metric dependent.


\bibliographystyle{IEEEtran}
\bibliography{cite}

\end{document}